\documentclass[useAMS,usenatbib]{mn2e}
\pdfoutput=1 
\usepackage{graphicx}    
\usepackage{dcolumn}     
\usepackage{bm}          
\usepackage{amssymb,amsmath}  
\usepackage{color}

\title[Precise Measurement of the Radial Baryon Acoustic Oscillation Scales in Galaxy Redshift Surveys]
      {Precise Measurement of the Radial Baryon Acoustic Oscillation Scales in Galaxy Redshift Surveys}
\author[E. S\'anchez et al.]
{E. S\'anchez$^1$\thanks{E-mail:eusebio.sanchez@ciemat.es}, 
 D. Alonso$^2$, F. J. S\'anchez$^1$, J.~Garc\'{\i}a-Bellido$^{2}$, I. Sevilla$^1$ \\
 $^1$Centro de Investigaciones Energ\'eticas, Medioambientales y Tecnol\'ogicas (CIEMAT), Madrid, Spain\\
 $^2$Instituto de F\'{\i}sica Te\'orica (UAM-CSIC), Madrid, Spain \\
}
\begin{document}

\date{\today}
\pagerange{1--11} \pubyear{2012}
\maketitle

\begin{abstract}
In this paper we present a new method to extract cosmological parameters using the radial 
scale of the Baryon Acoustic Oscillations as a standard ruler in deep galaxy surveys. The 
method consists in an empirical parametrization of the radial 2-point correlation 
function, which provides a robust and precise extraction of the sound horizon scale at the 
baryon drag epoch. Moreover, it uses data from galaxy surveys in a manner that is fully 
cosmology independent and therefore, unbiased. A study of the main systematic errors and 
the validation of the method in cosmological simulations are also presented, showing that 
the measurement is limited only by cosmic variance. We then study the full information 
contained in the Baryon Acoustic Oscillations, obtaining that the combination of the radial 
and angular determinations of this scale is a very sensitive probe of cosmological 
parameters, able to set strong constraints on the dark energy properties, even without 
combining it with any other probe. We compare the results obtained using this method with 
those from more traditional approaches, showing that the sensitivity to the cosmological 
parameters is of the same order, while the measurements use only observable quantities 
and are fully cosmology independent.
\end{abstract}

\section{Introduction}
\label{intro}

Finding the physical origin of the accelerated expansion of the Universe is one of the 
most important scientific problems of our time, and is driving important advances in
XXIst century cosmology. Several observational probes to study the nature of the mysterious
dark energy, which powers that expansion, have been proposed. Among them, the measurement 
of the scale of the Baryon Acoustic Oscillations (BAO) in the galaxy power spectrum as a 
function of redshift is one of the most robust, since it is insensitive to systematic 
uncertainties related to the astrophysical properties of the galaxies. Moreover, it 
provides information about dark energy from two different sources: the angular diameter 
distance, through the measurement of the BAO scale in the angular distribution of 
galaxies, and the Hubble parameter, through the measurement of the BAO scale in the 
radial distribution of galaxies.

There are some measurements of the BAO scale in the purely radial direction 
\citep{2009MNRAS.399.1663G, 2010ApJ...710.1444K, 2013MNRAS.431.2834X}, but most of them use 
mainly the monopole of the 3-D correlation function 
\citep{2005ApJ...633..560E, 2006A&A...449..891H,2006A&A...459..375H, 2007MNRAS.381.1053P,
2007MNRAS.378..852P, 2008ApJ...676..889O, 2009MNRAS.400.1643S, 2012MNRAS.427.3435A}. This 
has been the traditional way of determining the BAO scale, trying to optimize the sensitivity
when the number of galaxies in the survey is not very high, paying the price of introducing
model dependence in the measurement through the use of a fiducial model. However, the new 
galaxy surveys which are already taking data or those proposed for the future do not suffer 
from this problem and new and more robust methods can be used.

In this paper we propose a new method to extract the evolution of the radial BAO scale with 
the redshift, and explain how to use it as a standard ruler to determine cosmological 
parameters. We use data from galaxy surveys in a manner that is fully cosmology 
independent, since only observable quantities are used in the analysis and therefore, results 
are unbiased. A method based on the same idea for the measurement of the angular BAO scale 
was described in~\cite{2011MNRAS.411..277S} and provided the measurement of the angular 
BAO scale presented in \cite{2012MNRAS.419.1689C}. Here we present how to extract the 
radial BAO scale using a similar approach.

The method is designed to be used as a strict standard ruler, and provides the radial
BAO scale as a function of the redshift, but we do not intend to give a full description 
of the radial correlation function. This approach is more robust against systematic 
effects, and in fact we demonstrate that the measurement is only limited by cosmic 
variance, since the associated systematic errors are much smaller than the purely 
statistical errors.

\section{Galaxy Clustering and Observables}
\label{sec:theoryCorFunc}
One of the main statistical probes of the properties of the matter distribution in the
Universe is the 2-point correlation function, $\xi(r)$, which is defined as the excess 
joint probability that two point sources (\textit{e.g.} galaxies) are 
found in two volume elements $dV_1 dV_2$ separated by a distance $r$ compared to 
a homogeneous Poisson distribution~\citep{1980lssu.book.....P}. If the fluctuations on
the matter density field are Gaussian, this function contains all the information about 
the large scale structure of the Universe. 

However, what is observationally accesible is the distribution of galaxies in angle-redshift 
space, not directly the matter distribution in real space. For each galaxy we determine 
its angular position in the sky and its redshift. To obtain $\xi(r)$ we need to convert 
the measured redshift to a comoving distance, for which a cosmological model is 
needed. Therefore, the 3-D correlation function is not observable in a cosmology 
independent way for a given galaxy survey. Moreover, the observational techniques to 
obtain the angular position in the sky and the redshift are completely different and 
independent. Consequently, if we want to keep the measurement completely free of any 
theoretical interpretation, we should measure, on the one hand, the angular correlation 
function as a function of the angular separation of galaxy pairs, and on the other hand, the 
radial correlation function as a function of the redshift separation of galaxy pairs, and 
then extract the BAO scale from each function. Both of them are defined in complete analogy 
with the 3-D function, as the excess of probability with respect to a homogeneous Poisson 
distribution, and both are observable in a cosmology independent 
way, providing, together, information about the full 3-D large scale structure that can 
be described in any cosmological model. These aspects have been recently pointed out 
in \cite{2011PhRvD..84f3505B}, and here we present a practical implementation of a 
cosmology independent measurement, in this case for the radial BAO scale.

Several effects must be taken into account when the large scale structure of the Universe
is studied from galaxy surveys, like bias, redshift space distortions and non-linear 
corrections. In this paper we take all of them into account, both in the theoretical 
calculations and in the observational analysis. For this particular measurement, the 
influence of the non-linear effects is relatively small, since we are analyzing large 
scales, around the BAO scale. Also the influence of bias is small, because the analysis 
method is designed to minimize its impact on the final result. However, redshift space 
distortions are very important, since we are measuring along the line of sight.

\section{The Sound Horizon Scale}
\label{sec:soundhorizonscale}

The BAO are a consequence of the competition between gravitational attraction and gas 
pressure in the primordial plasma, which produced sound waves. Differences in density 
created by these sound waves leave a relic signal in the statistical distribution of 
matter in the Universe, defining a preferred scale, the sound horizon scale at the 
baryon drag epoch, which is a robust standard ruler from which the expansion history 
of the Universe can be inferred. The BAO scale is very large, $\sim 100$ Mpc/h, which 
poses an important challenge for observations, since surveys must cover large volumes 
to map such a distance. However, the very large scale of BAO also has certain 
advantages, because structure formation at these scales is rather well understood, and 
details about the description of galaxy formation and astrophysics do not compromise the 
accurate measurements of the standard ruler. In practice, the strength of the BAO standard 
ruler method relies on the potential to relate the position of the acoustic peak in the 
correlation function of galaxies to the sound horizon scale at decoupling.

The current measurements of the BAO scale typically use the spherically symmetric monopole 
contribution of the 3-D correlation function. This is a mixture of the angular and radial 
scales, and therefore, does not contain all the information that can be extracted from the 
data samples. On top of that, all analyses have been done using what is usually called a 
``fiducial model'' in order to determine distances. Rather than full tests of the 
cosmology, they should be understood as precision measurements within the context of the
standard $\Lambda$CDM cosmological model. Considering this way of analyzing data, what 
the current results have achieved is a strong confirmation of the consistency of the 
current data with the standard $\Lambda$CDM model, but the analysis strategy is 
very limited if one wants to test other non-FLRW cosmological models against data, or at 
least, the whole analysis must be repeated from the very begining for every cosmological 
model one tries to test.

On the other hand, if both the angular and the radial BAO scales are determined, which in 
principle is already possible with current data sets, the cosmological constraints that can 
be set using only the BAO scale are much stronger than the current values, since the 
radial and angular scales are sensitive to cosmological parameters in quite different 
ways, breaking degeneracies and allowing a very precise determination when combined. 

\section{Method to Measure the Radial BAO Scale}
\label{sec:method}

The method we propose relies on a parametrization of the radial correlation function
which allows extracting the radial BAO scale with precision. It is based on a similar 
approach previously proposed for the angular correlation 
function~\citep{2011MNRAS.411..277S}. The suggested parametrization was inspired by 
the application of the Kaiser effect to the shape we used for the angular correlation 
function. 

It is important to remark that the calculations we perform to check the goodness of the 
parametrization do include redshift space distortions using the Kaiser description 
\citep{1987MNRAS.227....1K} and also non-linearities using the RPT approach 
\citep{2006PhRvD..73f3519C}, and we expect the result to be correct for large 
scales, where the neglected effects (fingers of god, mode-mode coupling) are very 
small. We have used a bias parameter $b=1$ throughout the calculation. The possible 
influence of bias is studied in the systematic errors section. We use the linear power 
spectrum from {\tt CAMB} \citep{CAMB}, on top of which non-linearities and redshift 
space distortions are then included.

\begin{figure*}
  \centering
  \leavevmode
  \begin{tabular}{ccc}
    \includegraphics[width=0.31\textwidth]{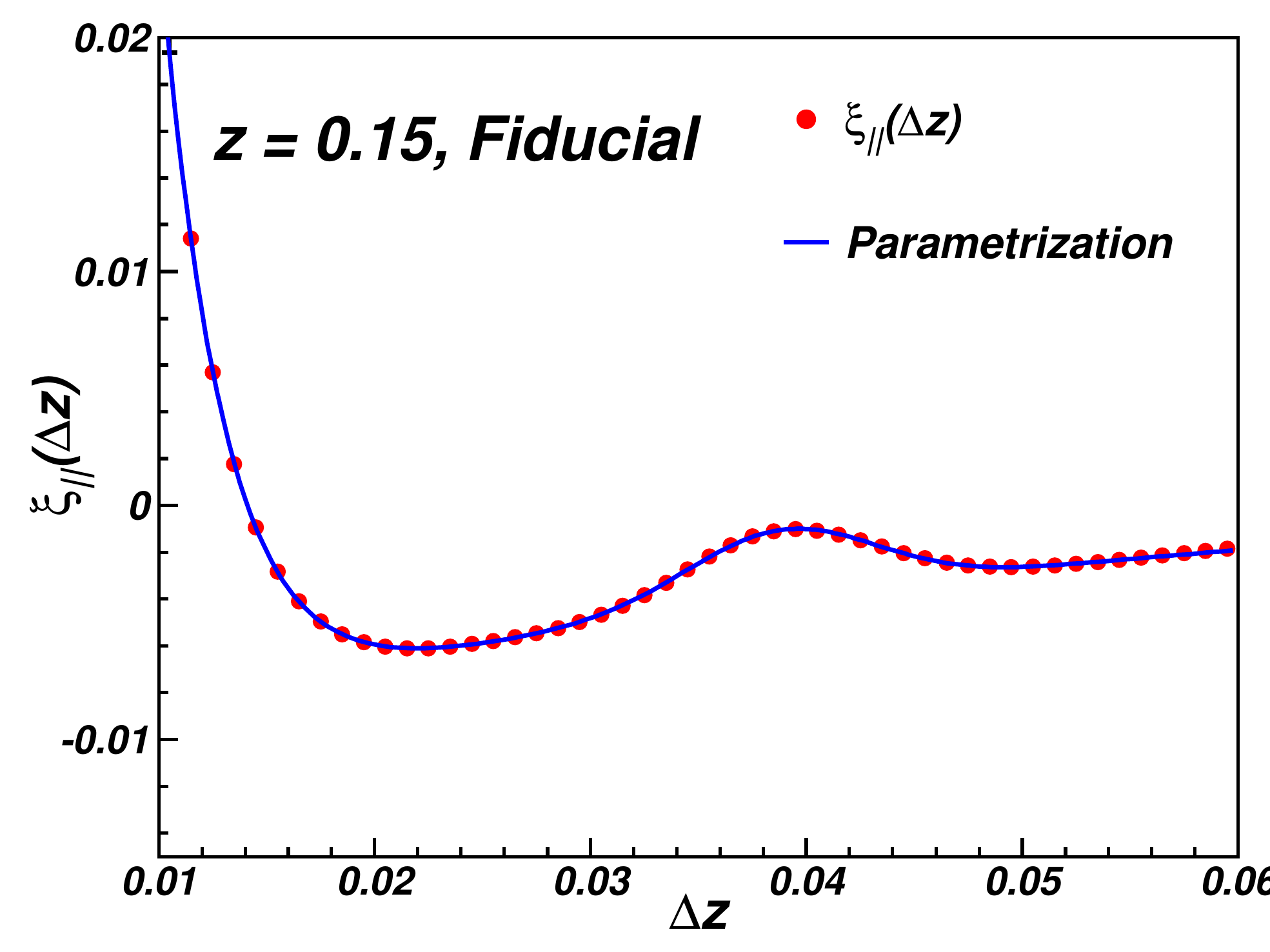}  &
    \includegraphics[width=0.31\textwidth]{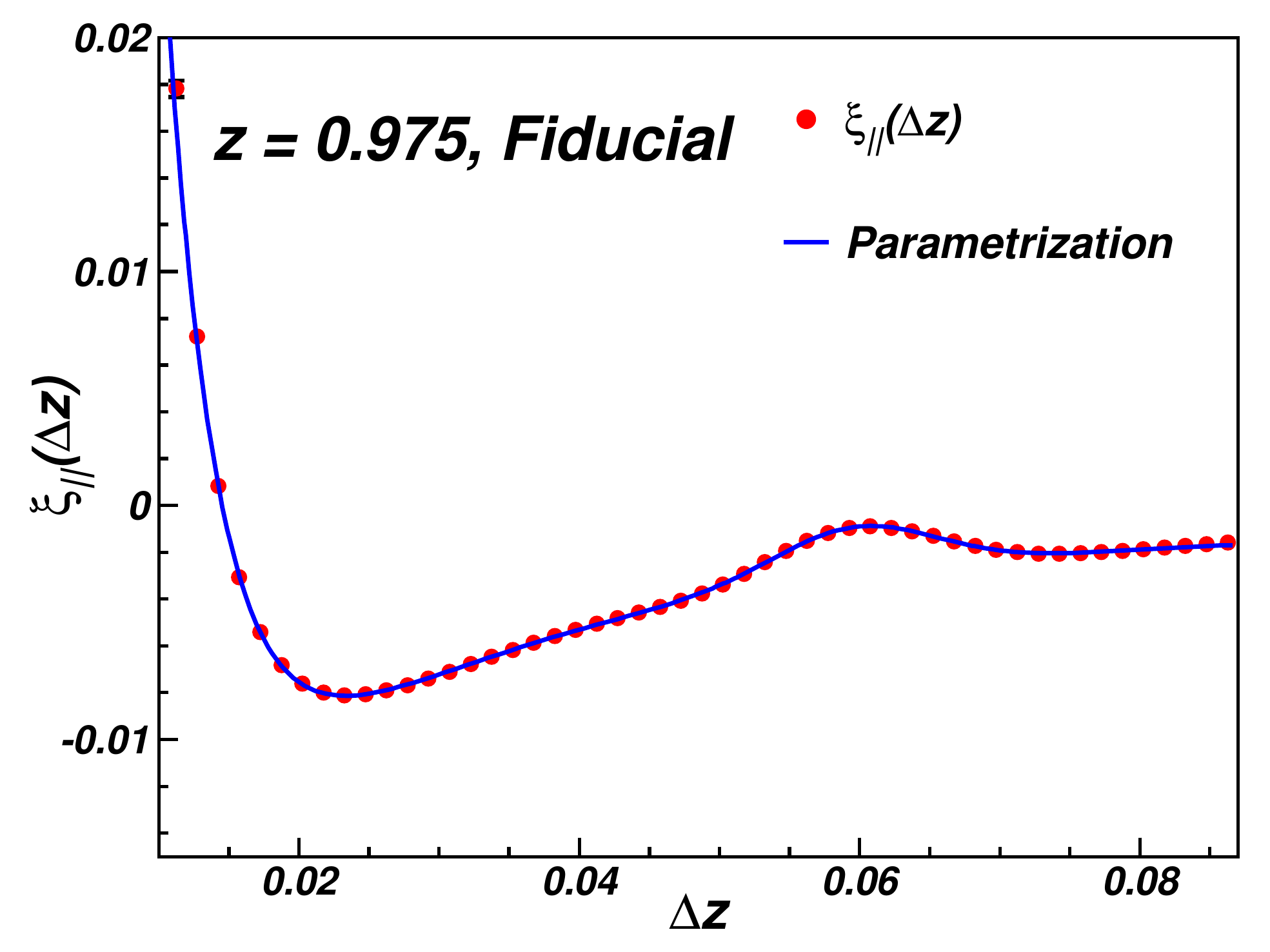} &
    \includegraphics[width=0.31\textwidth]{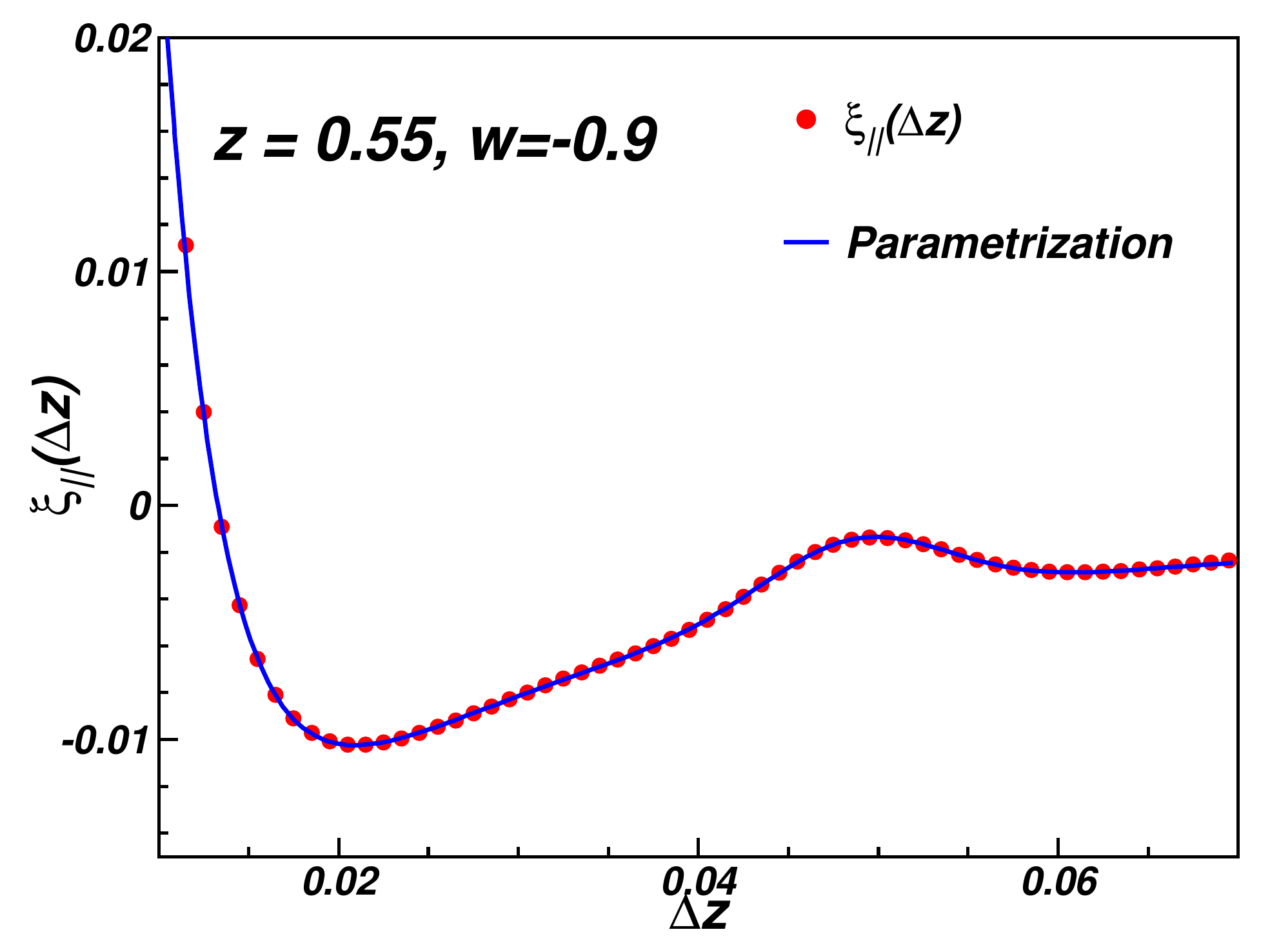}  \\
    \includegraphics[width=0.31\textwidth]{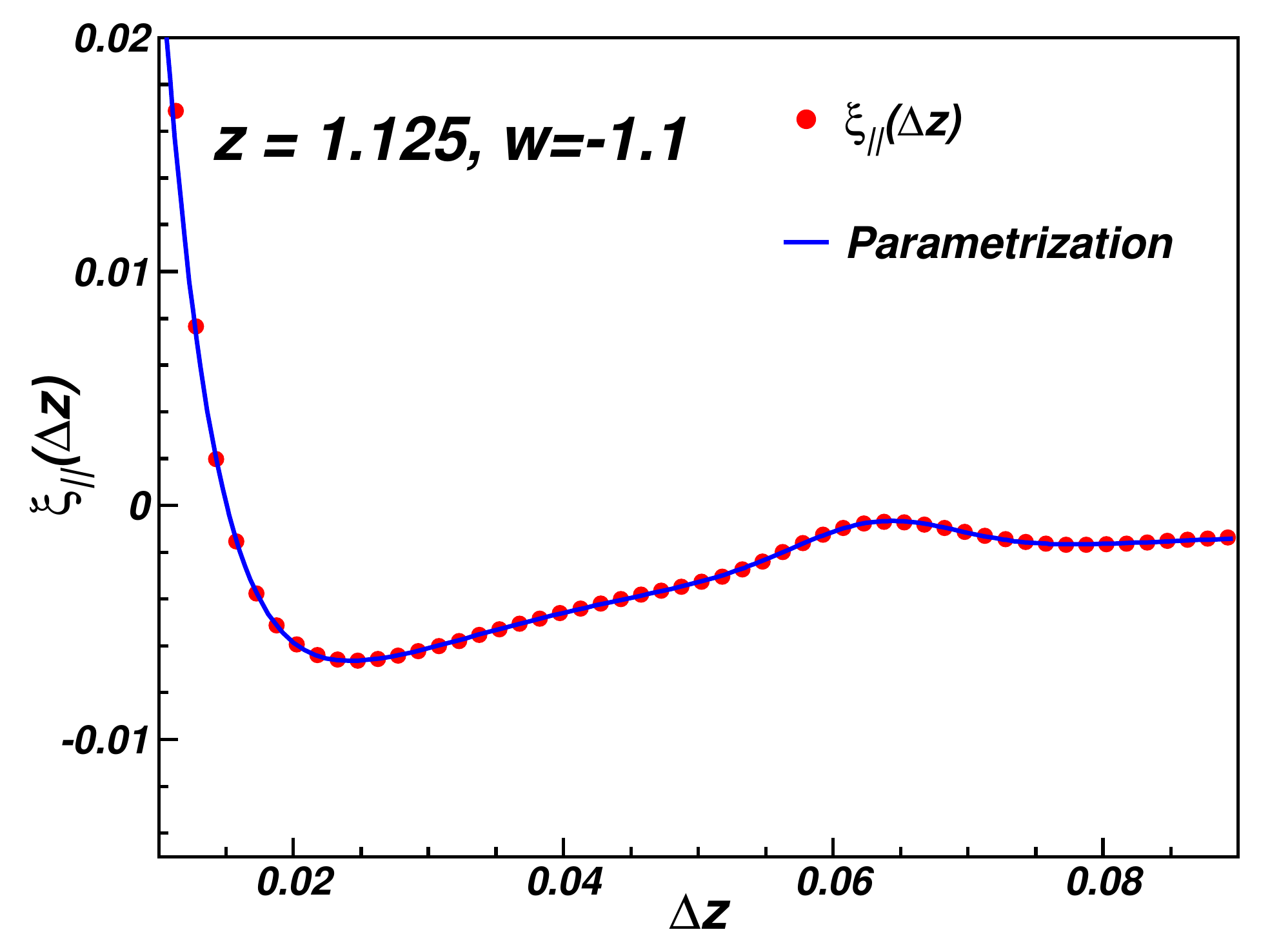} &
    \includegraphics[width=0.31\textwidth]{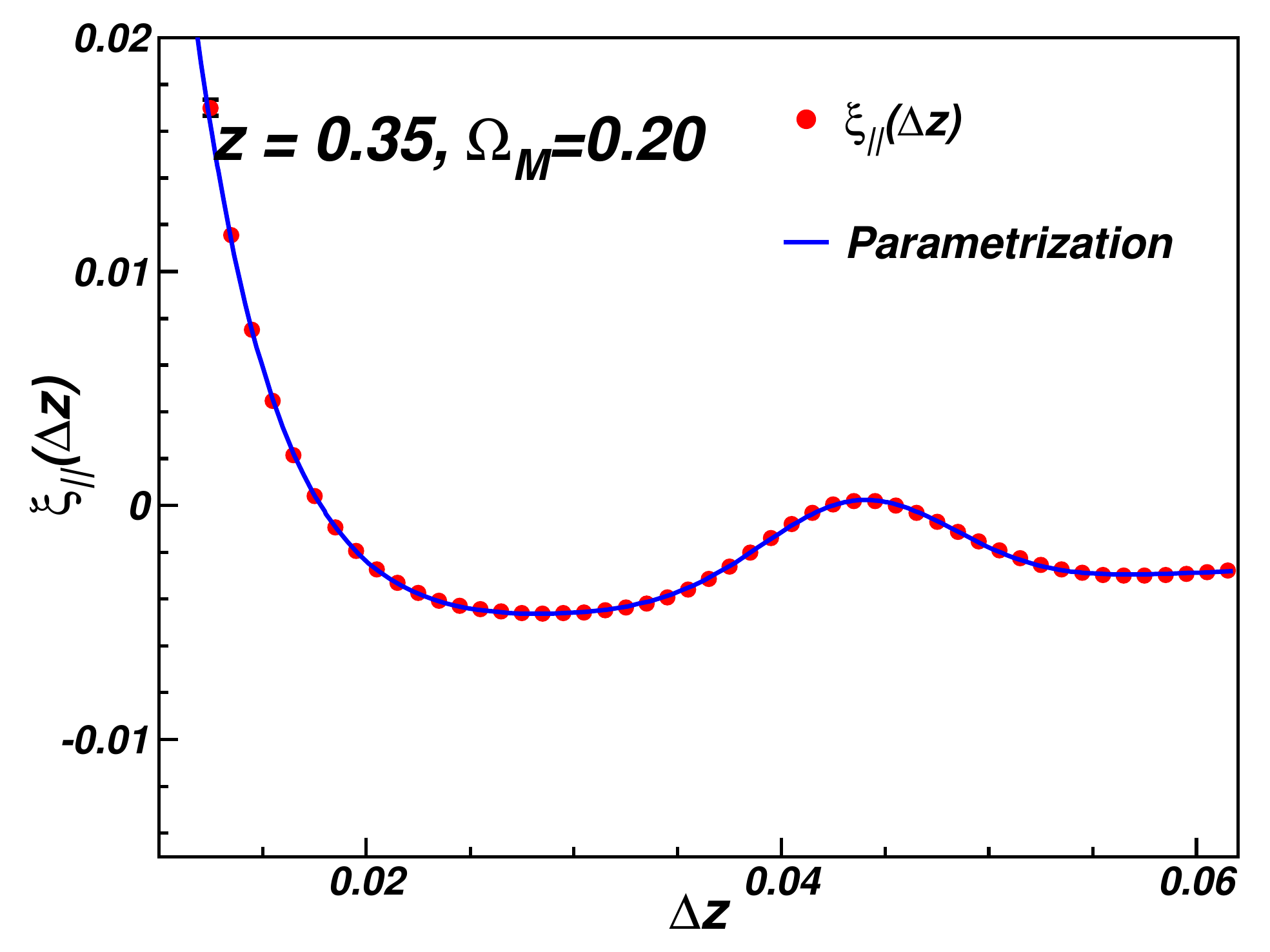}  &
    \includegraphics[width=0.31\textwidth]{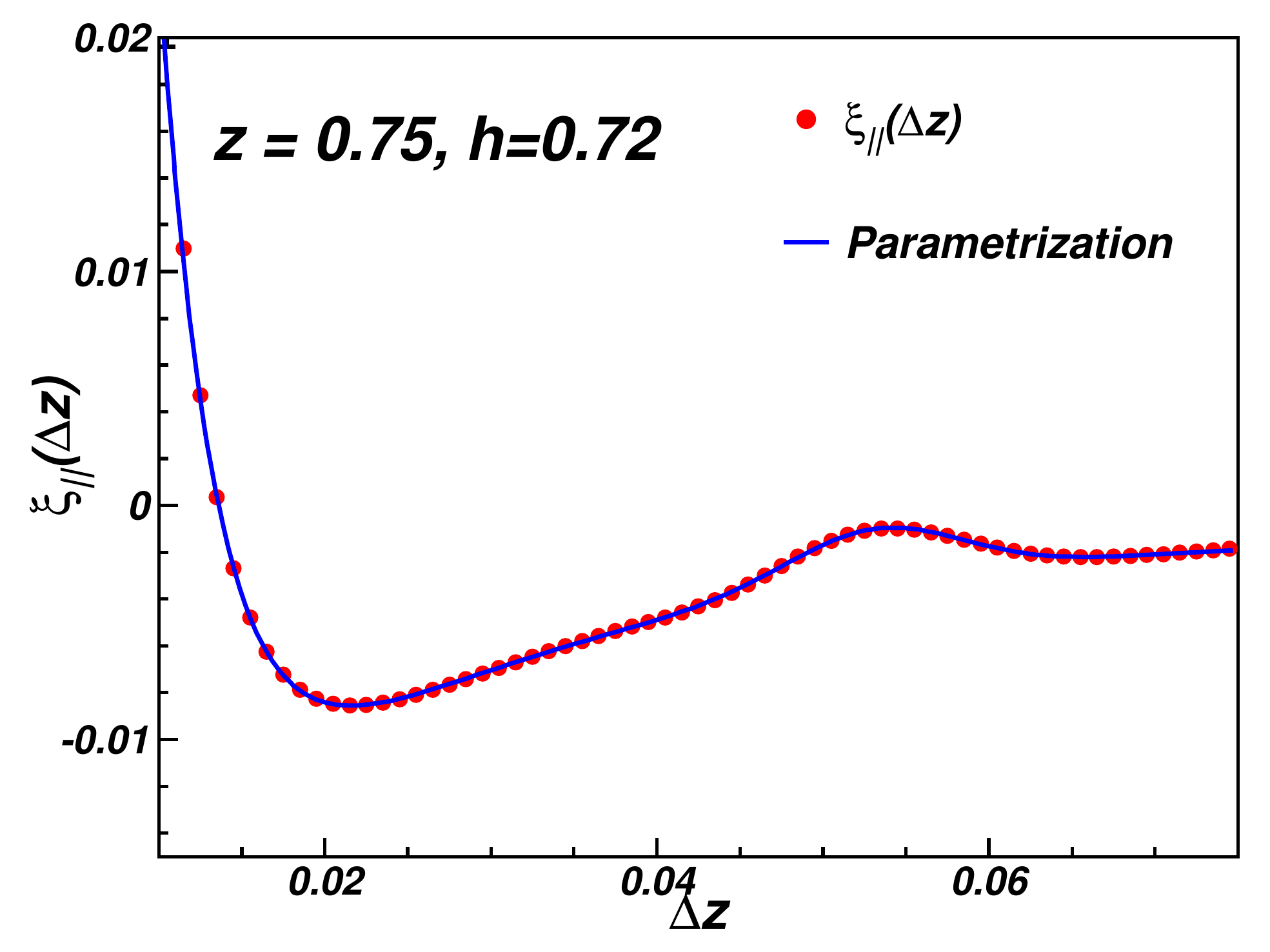}  \\
  \end{tabular}
  \caption{Some examples of the description of the different cosmological models 
           by the proposed parametrization. Models do include redshift space distortions and
           non-linearities using the RPT scheme. The parametrization is a very good
           description of all models at all redshifts, even if the considered errors
           are much smaller than the cosmic variance for the full sky. Errors are smaller
           than the dot size. We have used this level of precision to ensure that the 
           systematic errors associated to the theory (non-linearities, bias) are small.}
  \label{fig:theopar}
\end{figure*}

The full recipe to obtain the radial BAO scale as a function of redshift for a galaxy survey is as 
follows:

\begin{enumerate}
\item Divide the full galaxy sample in redshift bins.

\item Divide each redshift bin in angular pixels.

\item Compute the radial correlation function by stacking galaxy pairs in each angular pixel, but
do not mix galaxies in different angular pixels.

\item Parametrize the correlation function using the expression:
\begin{equation}
\label{eq:param}
\xi_{\parallel}(\Delta z) = A + B e^{-C \Delta z} - D e^{-E \Delta z} 
                          + F e^{-\frac{(\Delta z - \Delta z_{BAO})^{2}}{2\sigma^2}}
\end{equation} 

\noindent
and perform a fit to $\xi_{\parallel}(\Delta z)$ with free parameters 
$A$, $B$, $C$, $D$, $E$, $F$, $\Delta z_{BAO}$ and $\sigma$. 

\item The radial BAO scale is given by the parameter $\Delta z_{BAO}$. The BAO scale as 
a function of the redshift is the only parameter needed to apply the standard ruler 
method. The cosmological interpretation of the other parameters is limited, since this is 
an empirical description, valid only in the neighborhood of the BAO peak.

\item Fit cosmological parameters to the evolution of $\Delta z_{BAO}$ with $z$.
\end{enumerate}

In order to test the goodness of this parametrization, we have computed the radial 
correlation function for the 14 cosmological models described in Table~\ref{tab:cosmomodels} in
a redshift range from 0.2 to 1.5, always including redshift space distortions and
non-linearities as explained before. Then, we have applied the method to each model and each 
redshift. The parametrization describes the theory very accurately, since the
values of the $\chi^2/ndof$ are close to 1, and the probabilities of the fit lie between 
0.98 and 1.00 when the error in each point of the correlation function is arbitrarily 
fixed to 2\% (or $\sim 10^{-5}$) for all bin widths and cosmological models. This error 
corresponds to a precision much better than the cosmic variance for the full sky, for 
all models in the full redshift range, but the parametrization is able to recover the 
correct radial BAO scale for the 14 cosmologies. We have used this level of precision 
because the systematic errors coming from theoretical effects (non-linearities, bias, fingers 
of god) only affect this method if they induce a disagreement between the models and 
the parametrization, giving then a wrong measurement of the BAO scale. If the description 
is good to this level of precision we guarantee a small contribution from these systematic 
effects, as will be shown in the systematic errors section below. The calculation of 
the realistic errors is described in the next section. Some examples of these descriptions 
can be seen in Figure~\ref{fig:theopar}.

Once we have verified the parametrization on the theoretical predictions, we are ready to
apply it in more realistic environments. For this purpose, in the next section we apply 
this algorithm to a galaxy catalogue obtained from a N-body cosmological simulation. We will
also compute the main systematic errors associated to this method.

\begin{table}
\centering
\begin{tabular}{ccccccc}
\hline
$h$ & $\Omega_{M}$ & $\Omega_{b}$ & $\Omega_{k}$ & $w_{0}$ & $w_{a}$ & $n_{s}$ \\
\hline\hline
0.70 & 0.25 & 0.044 &  0.00   & $-$1.00 &  0.0   & 0.95 \\
0.68 & ~    & ~     & ~       & ~       & ~      & ~    \\
0.72 & ~    & ~     & ~       & ~       & ~      & ~    \\
~    & 0.20 & ~     & ~       & ~       & ~      & ~    \\ 
~    & 0.30 & ~     & ~       & ~       & ~      & ~    \\ 
~    & ~    & 0.040 & ~       & ~       & ~      & ~    \\
~    & ~    & 0.048 & ~       & ~       & ~      & ~    \\
~    & ~    & ~     & +0.01   & ~       & ~      & ~    \\
~    & ~    & ~     & $-$0.01 & ~       & ~      & ~    \\
~    & ~    & ~     & ~       & $-$0.90 & ~      & ~    \\
~    & ~    & ~     & ~       & $-$1.10 & ~      & ~    \\
~    & ~    & ~     & ~       & ~       & $-$0.1 & ~    \\
~    & ~    & ~     & ~       & ~       & +0.1   & ~    \\
~    & ~    & ~     & ~       & ~       & ~      & 1.00 \\
\hline
\end{tabular}
\caption{Summary of the 14 cosmological models used to test the method.  Where empty, the 
fiducial values (first line) are assumed.\label{tab:cosmomodels}}
\end{table}

\section{Application to a simulated galaxy Survey}
\label{sec:simulation}

\begin{figure}
  \centering
  \includegraphics[width=0.49\textwidth]{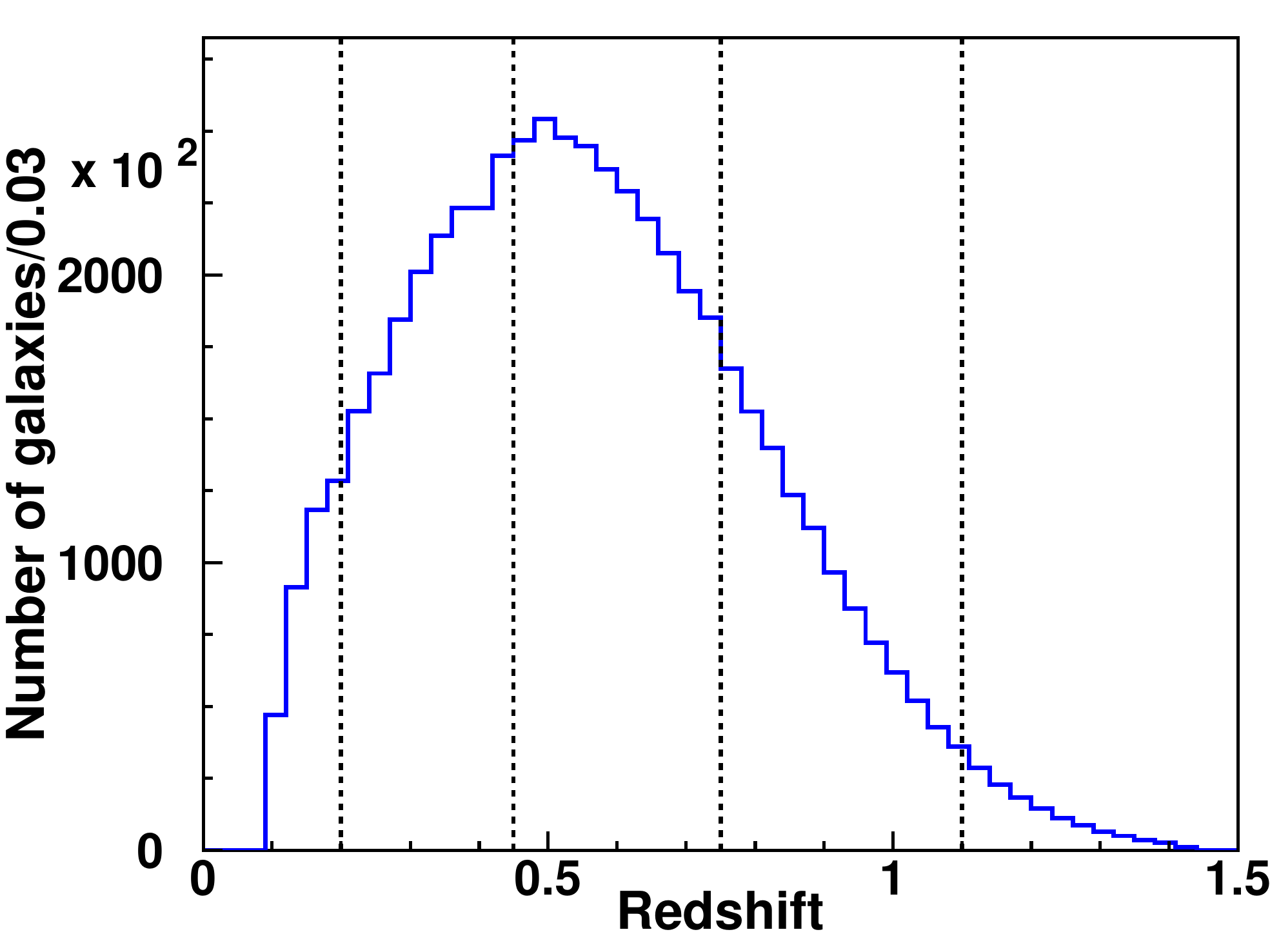}
  \caption{$N(z)$ in the used MICE catalogue. The simulation contains 55 million 
           galaxies in the redshift range $0.1 < z < 1.5$. The vertical dashed lines
           show the limits of the redshift bins used in the analysis.}
  \label{fig:dndzMICE}
\end{figure}

We have tested the method to recover the radial BAO scale using a large N-body
simulation capable of reproducing the geometry (e.g. area, density and depth) and
general features of a large galaxy survey. The simulated data were kindly provided 
by the MICE project team, and consisted of a distribution of dark matter 
particles (galaxies, from now on) with the cosmological parameters fixed to the
fiducial model of Table \ref{tab:cosmomodels}. The redshift distribution of the galaxies
is shown in Figure \ref{fig:dndzMICE}. The simulation covers 1/8 of the full sky (around 
5000 square degrees) in the redshift range 0.1 $<$ z $<$ 1.5, and contains 55 million 
galaxies in the lightcone. This data was obtained from one of the largest N-body simulations 
completed to date \footnote{http://www.ice.cat/mice}, with 
comoving size $L_{box} = 3072~ h^{-1}$ Mpc and more than $8 \times 10^{9}$ particles 
($m_p = 2.3 \times 10^{11} h^{-1} M_{\odot}$). More details about this simulation can be found 
in \cite{2008MNRAS.391..435F} and \cite{2010MNRAS.403.1353C}. The simulated catalogue contains 
the effect of the redshift space distortions, fundamental for the study of the radial BAO scale.

Data with similar characteristics will be obtained in future large spectroscopic 
surveys, such as DESpec \citep{2012arXiv1209.2451A}, BigBOSS \citep{2011arXiv1106.1706S} or
{\it EUCLID} \citep{2011arXiv1110.3193L}.

It is important to note that the radial BAO determination needs a very large survey 
volume. We tried to extract the BAO peak from catalogs with smaller areas (200, 500 
and 1000 sq-deg), finding a very small significance (or no detection at all) in most 
cases. This is due to the fact that the statistical error related to the cosmic variance 
is specially large for the radial correlation function, and therefore it can only be 
reduced by increasing the volume explored. 

\begin{figure}
  \centering
  \begin{tabular}{c}
    \includegraphics[width=0.40\textwidth]{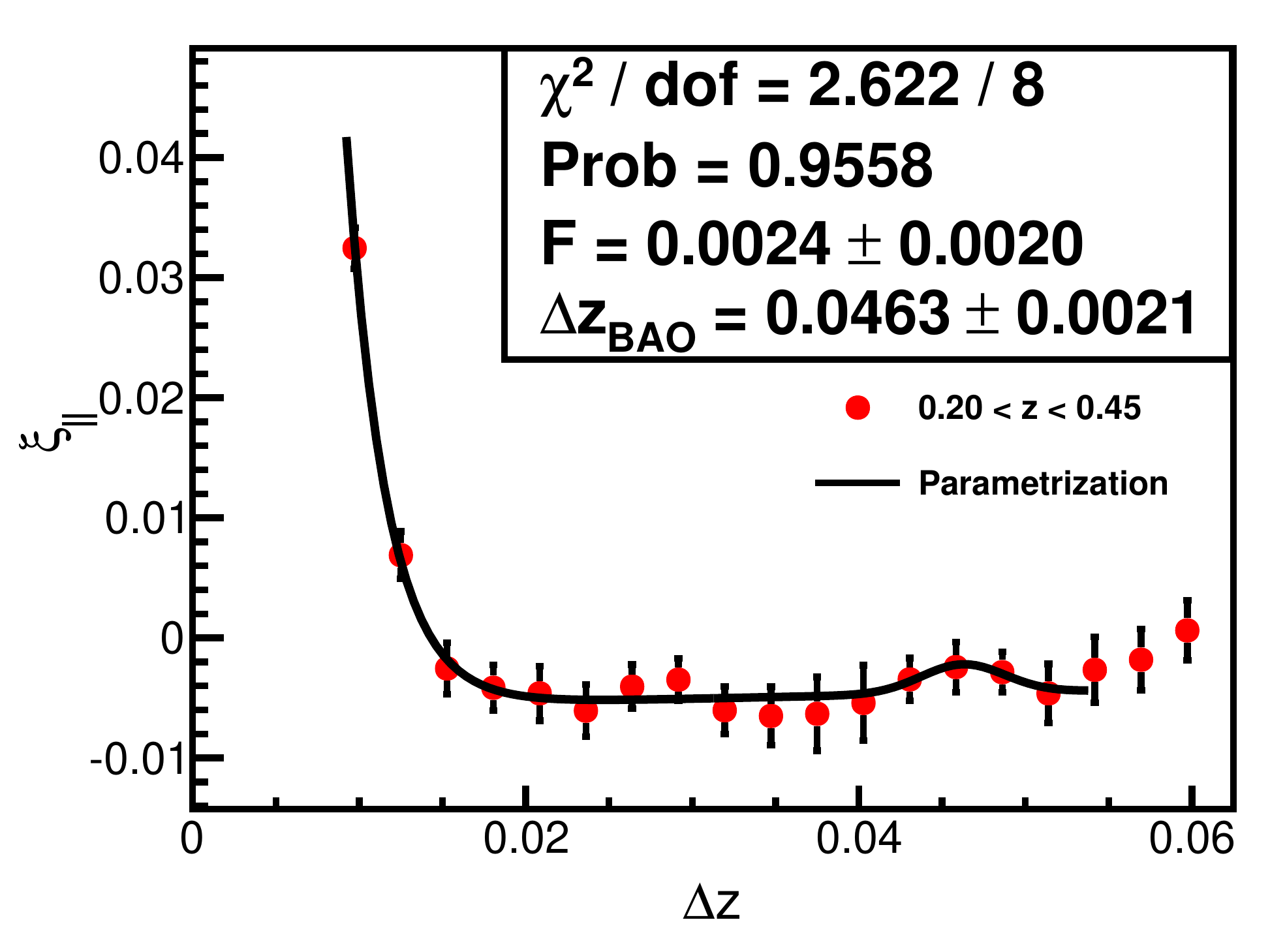} \\
    \includegraphics[width=0.40\textwidth]{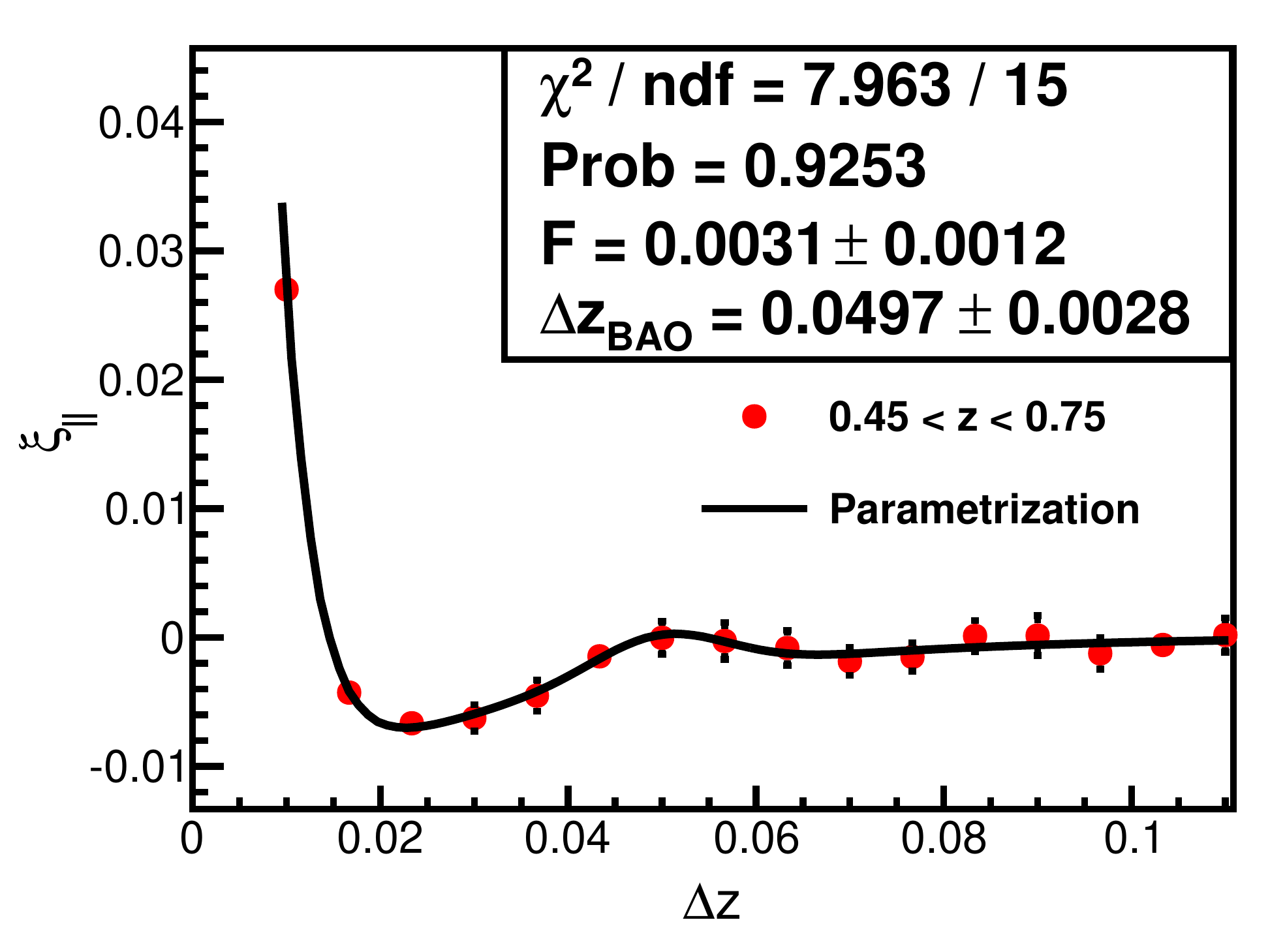} \\
    \includegraphics[width=0.40\textwidth]{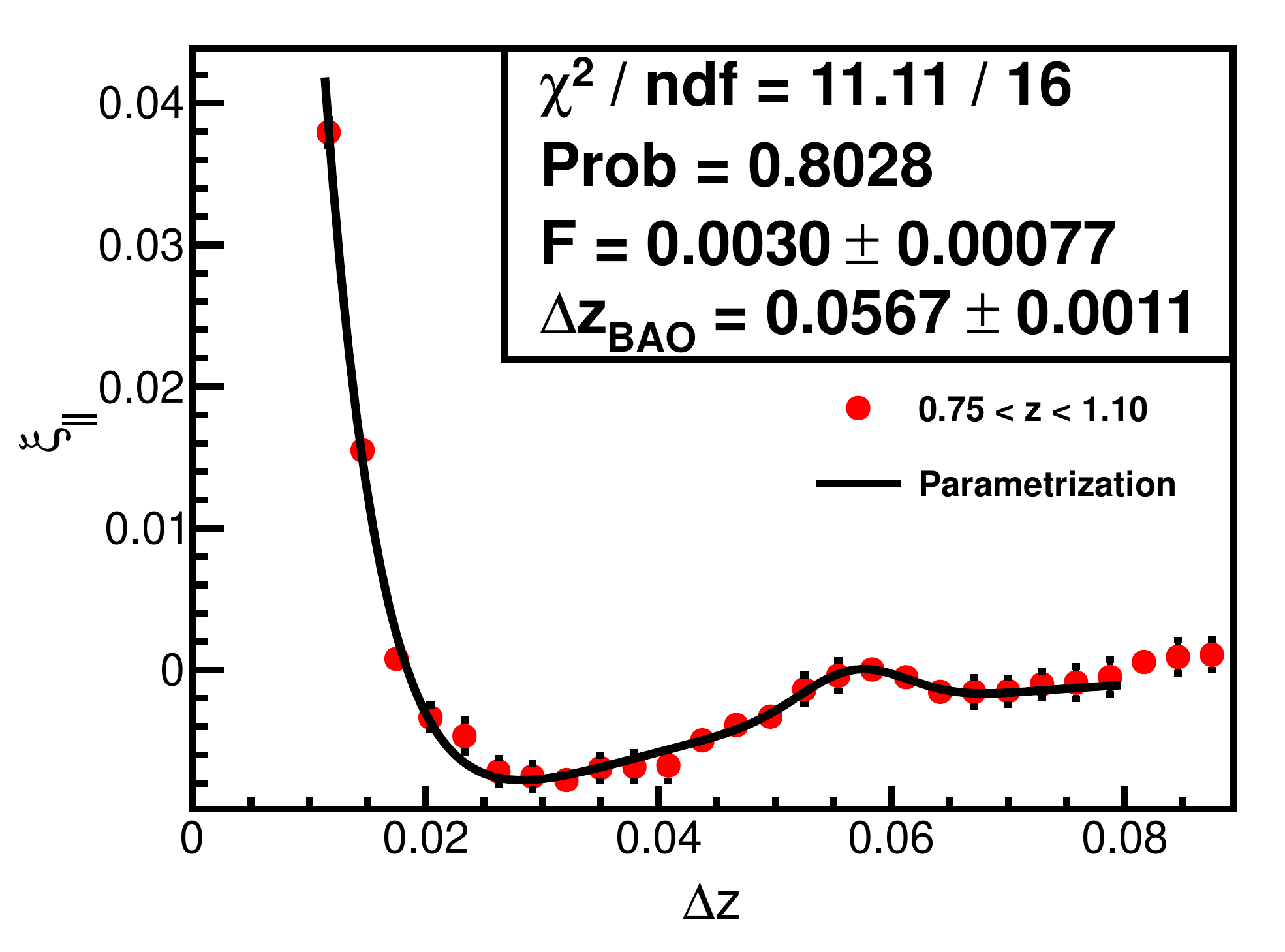} \\
    \includegraphics[width=0.40\textwidth]{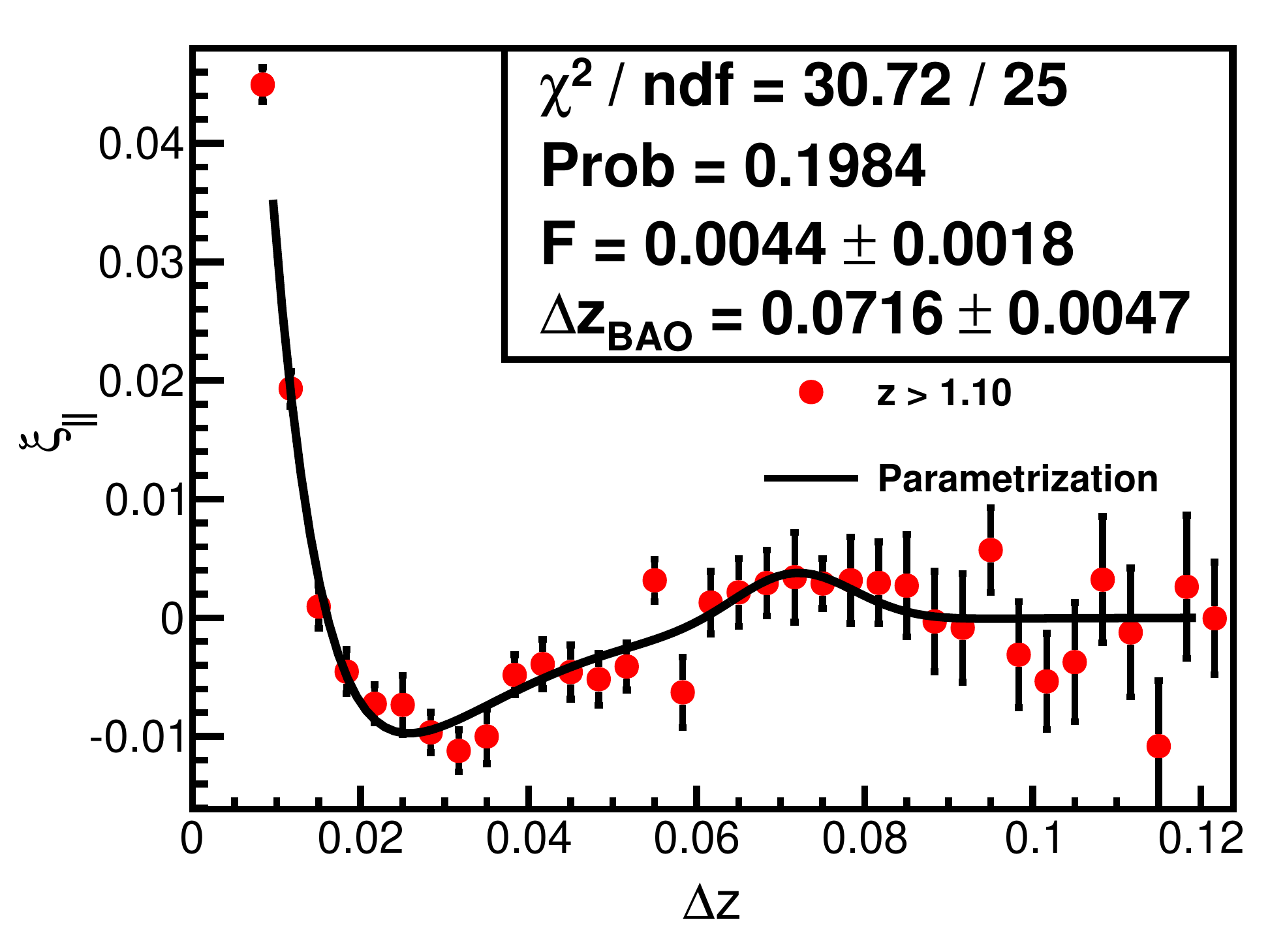} \\
  \end{tabular}
  \caption{Radial correlation functions measured in the MICE simulation for the 4 redshift
           bins described in the text (dots), for an angular pixel of 0.25 sq-deg, compared 
           with the proposed parametrization (solid line). All the fits are good. The statistical 
           significance of the BAO detection in the first bin is very low, and it is not used to 
           set cosmological constraints.}
  \label{fig:micetruez}
\end{figure}

Thus, we have divided the simulation into 4 redshift bins. We apply the method 
described in the previous section for each bin and obtain the correlation functions 
using the Landy-Szalay estimator \citep{1993ApJ...412...64L}. Fits are shown in 
Figure~\ref{fig:micetruez}. We have used an angular pixel with of 
02.5 sq-deg size, in order to retain enough number of galaxy pairs in the colliner
direction. We generate the random catalogues
taking into account the $N(z)$ distribution in order to obtain the correct determination 
of the radial correlation function. This is a very important effect, since we are 
measuring in the radial direction and the distributions on the redshift coordinate 
have a large influence in measurements. The statistical significance of the BAO 
observation in the first bin is very low ($\sim 1.4 \sigma$) and it is consequently 
not considered in the cosmology analysis. Following the same approach 
of \cite{2011MNRAS.411..277S}, we have computed the statistical significance of the 
detection by measuring how different from zero the $F$ parameter of the 
fit is Eq.~(\ref{eq:param}), using its statistical error.

\begin{figure}
  \centering
  \includegraphics[width=0.45\textwidth]{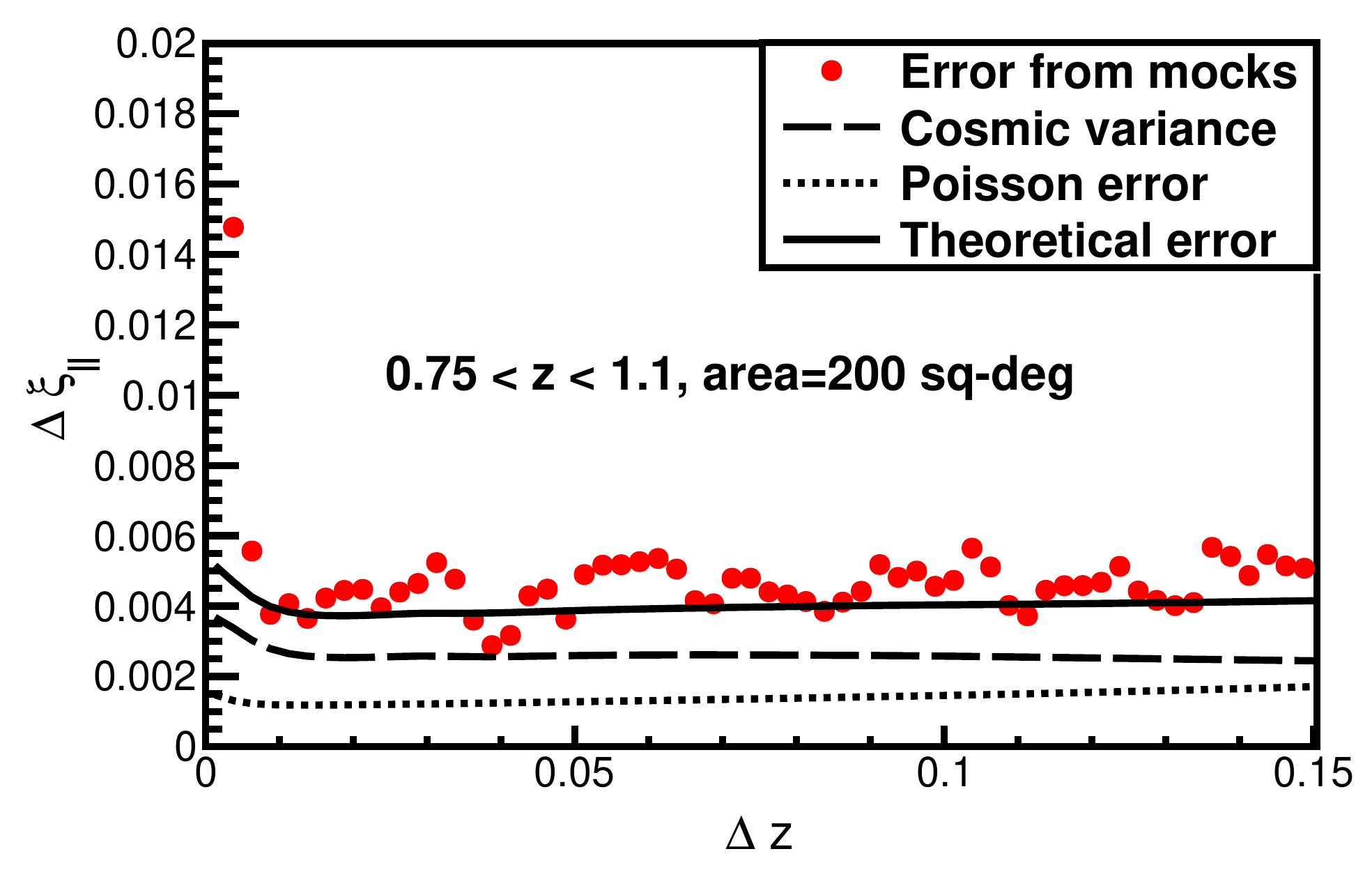}
  \caption{Comparison of the different estimates of the error in the radial
           correlation function are shown. The estimates are MC samples (dots) and 
           theoretical calculation (solid line). They agree in the region of interest for
           the BAO analysis. The disagreement at low scales comes from the incomplete 
           description of the non-linearities, where the mode-mode coupling effect is
           neglected, and from some boundary effects in the realizations, but
           does not affect the BAO scale measurement since it is outside the fitting region. The 
           different contributions to the total error are shown. They come from Poisson shot 
           noise (dotted line) and cosmic variance (dashed line). The contribution of the 
           Poisson shot noise, proportional to the inverse of the pair counts, is not negligible.}
  \label{fig:XiError}
\end{figure}

We have computed a theoretical estimation of the covariance matrix for the 3-D correlation
function (see the Appendix \ref{sec:error3-D}), which is then applied to the estimation of 
the covariance for the radial correlation function Eq. (\ref{eq:gcv_radial}), taking the 
corresponding value for the line of sight direction. This estimate relies on the model 
used for the simulation, but we expect a small variation of the error with the cosmological 
model. In any case, to keep the method fully model independent we have validated the 
calculation of the covariance matrix obtaining it with an alternative method directly from 
the simulation, which can also be applied to any real catalogue. We have used many 
realizations, dividing the total area in patches of different sizes and computing the 
corresponding dispersions and covariance matrices. We have then obtained the error for 
the total area, scaling this estimates to the full area of the measurement. Both 
determinations agree in the region of interest for the BAO scale measurement, as is 
shown in Figure \ref{fig:XiError}. There is a disagreement for small scales, which
is coming from the incomplete description of the non-linearities in the theoretical 
calculation, where the mode-mode coupling effects are neglected, and from boundary effects
in the realizations.

It is important to note that the radial BAO determination needs a very large survey 
volume. We are not able to obtain a significant observation for the first redshift bin 
and the contribution to the total error of the cosmic variance and the shot noise is 
comparable for all the other bins, which is shown in Figure \ref{fig:XiError}. As 
already pointed out and verified by \cite{2011MNRAS.414..329C}, in order to obtain a 
correct description of the error we need to sum linearly the cosmic variance and the 
shot noise contribution in harmonic space. This is not surprising, since galaxy 
surveys are designed to have a determined and fixed galaxy density, and this fact 
correlates both contributions to the error.

\begin{figure}
  \includegraphics[width=0.50\textwidth]{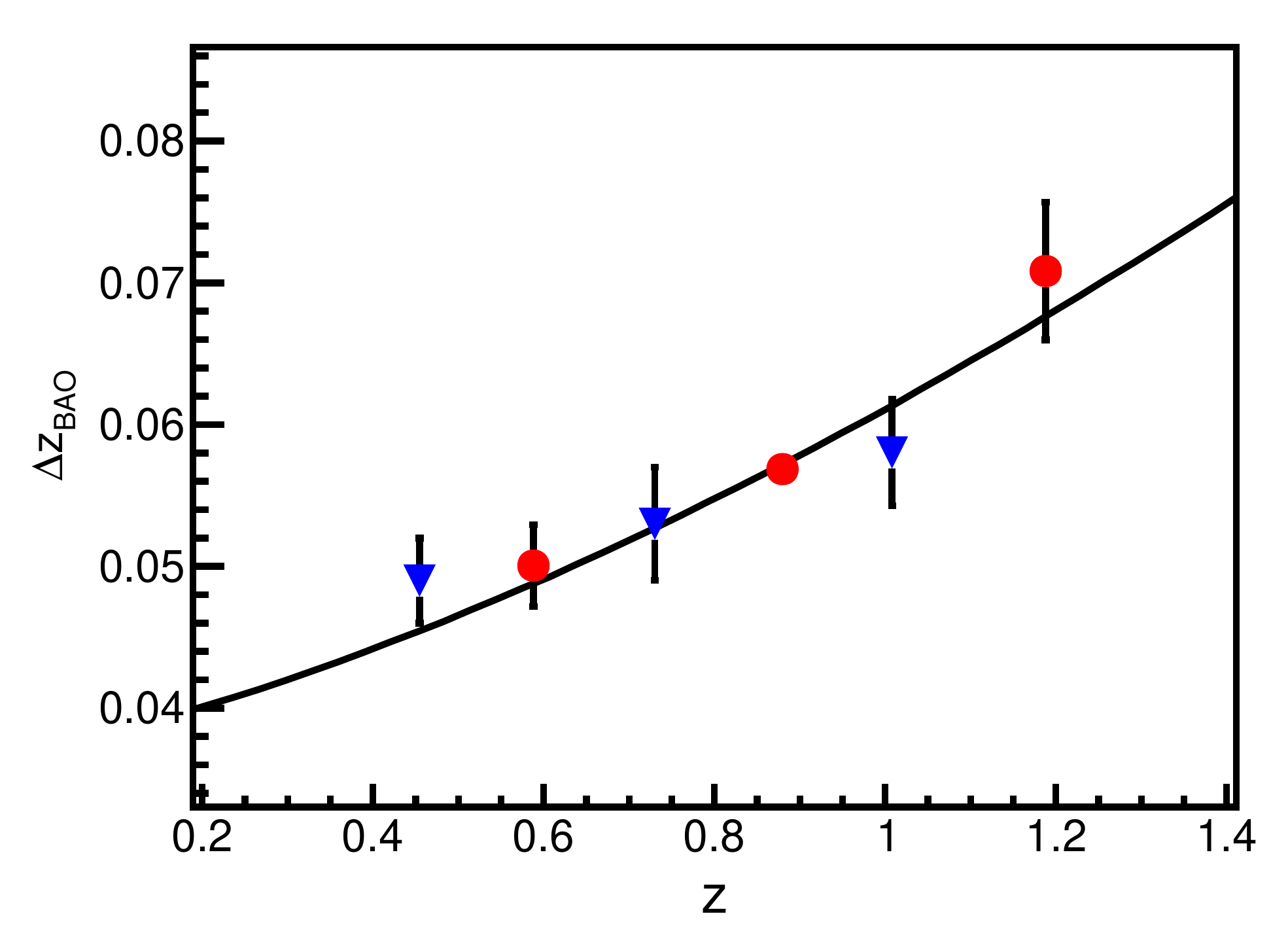}
  \caption{Measured radial BAO scale as a function of the redshift in the MICE simulation
           using the proposed method. Dots are the nominal bins
           and triangles correspond to displaced bins, and are measured only as a cross-check. All
           measurements are in good agreement with the theoretical prediction (solid line).}
  \label{fig:dzBAOvsz}
\end{figure}

The measured values of the BAO scale as a function of the redshift can be seen in 
Figure~\ref{fig:dzBAOvsz} as dots. The results for 3 alternative bins shifted with 
respect to the nominal ones are also shown as triangles. These have not
been used to obtain cosmology results, since they are fully correlated with the
red ones and are only shown for illustration and verification of the correct behaviour of
the analysis method. We have used in this analysis the center of the redshift bin to obtain the
prediction of the model, although what is really observed is the average within the bin. We have 
verified that they are very close if the $N(z)$ distribution is smooth, as it is in this case. 

\begin{figure}
  \includegraphics[width=0.45\textwidth]{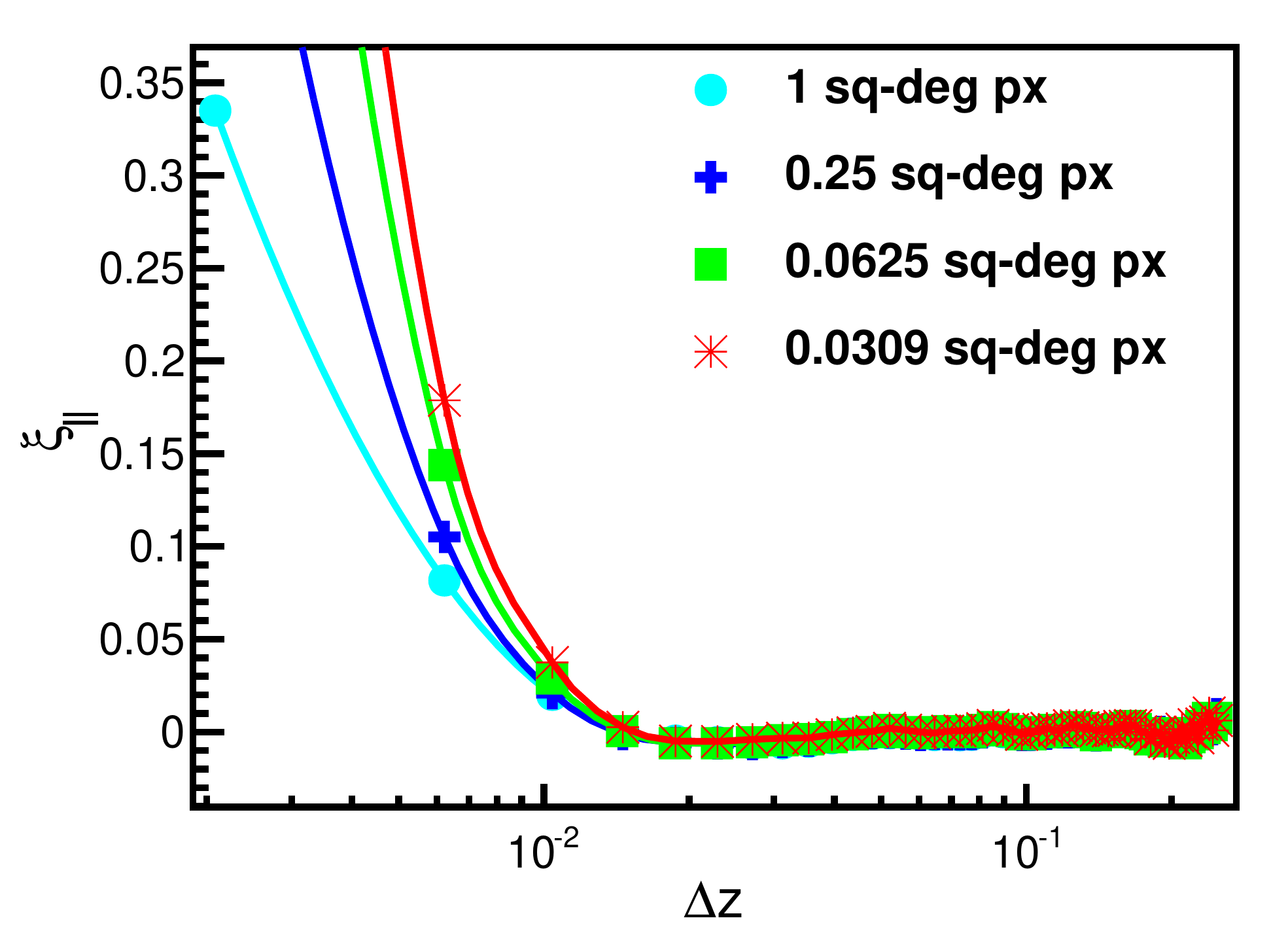}
  \includegraphics[width=0.45\textwidth]{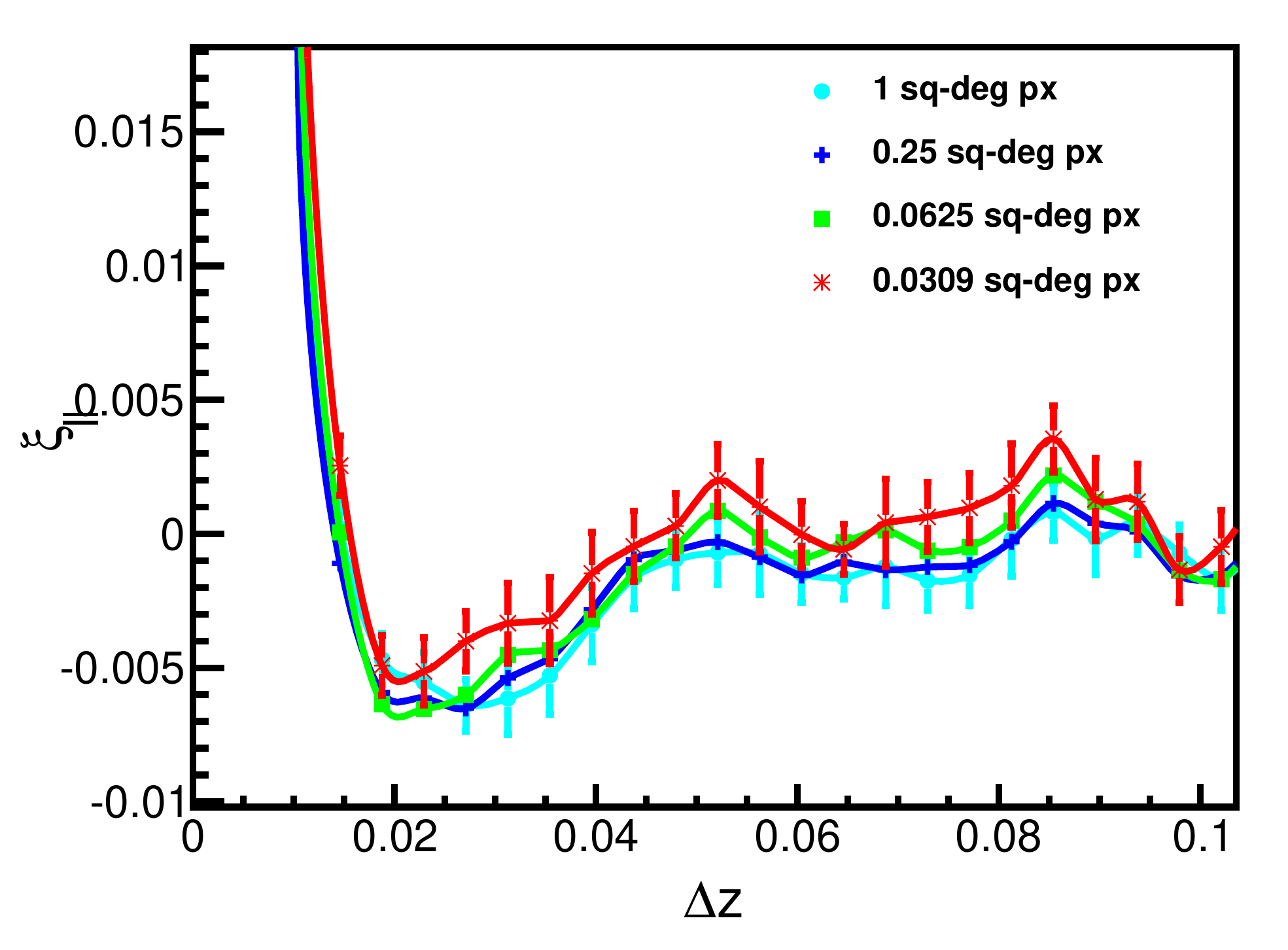}
  \caption{Effect of the angular pixel size on the radial correlation function. The finite 
           size of the angular pixel induces a change on the slope of the correaltion function 
           at small scales (top), but when a zoom around the position of the BAO peak is done, it
           is clear that this effect does not change the position of the BAO peak (bottom). The 
           effect is shown for pixels of sizes 0.0309 (stars), 0.0625 (squares), 0.25 (crosses) 
           and 1 (dots) square degrees. The 0.25 sq-deg pixel has been used to obtain the cosmological
           parameters. The change in the slope arises from the smoothing effect produced 
           by the inclusion in the calculation of galaxy pairs which are not exactly collinear. This 
           effect does not affect the determination of the BAO scale.}
  \label{fig:pixelsize}
\end{figure}

\subsection{Systematic Errors}
\label{sec:sys}

We have studied the main systematic errors that affect the determination of the radial BAO scale using
this method. We have found a specific systematic effect associated to the algorithm, which is 
coming from the size of the angular pixel. Other systematic errors are generic and will be present 
in any determination of the BAO scale, namely, the influence of the non-linearities, the starting 
and end point of the fit to the correlation function and the possible influence of the galaxy 
bias in the measurement. 

\begin{figure}
  \begin{tabular}{c}
    \includegraphics[width=0.40\textwidth]{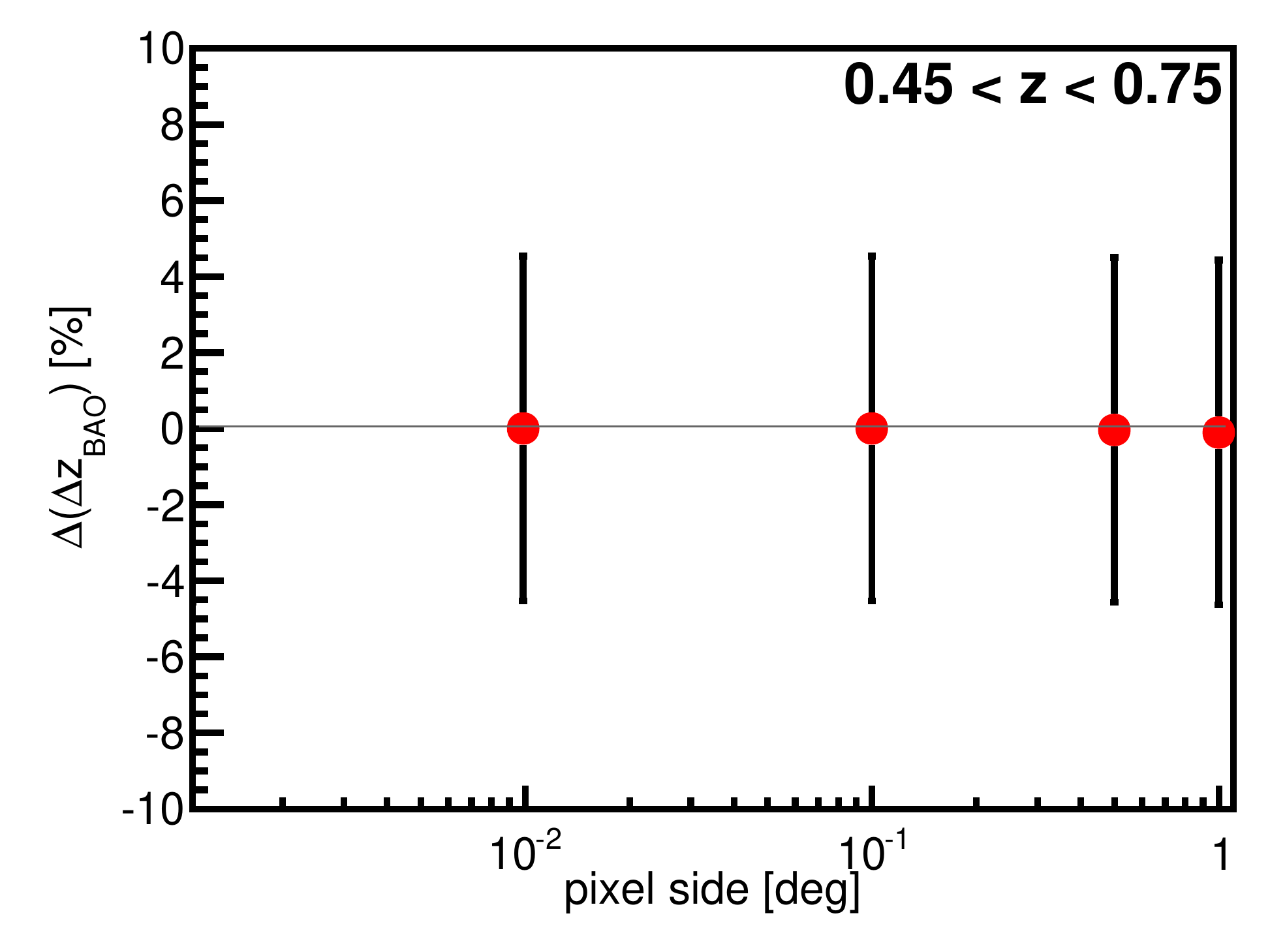}  \\
    \includegraphics[width=0.40\textwidth]{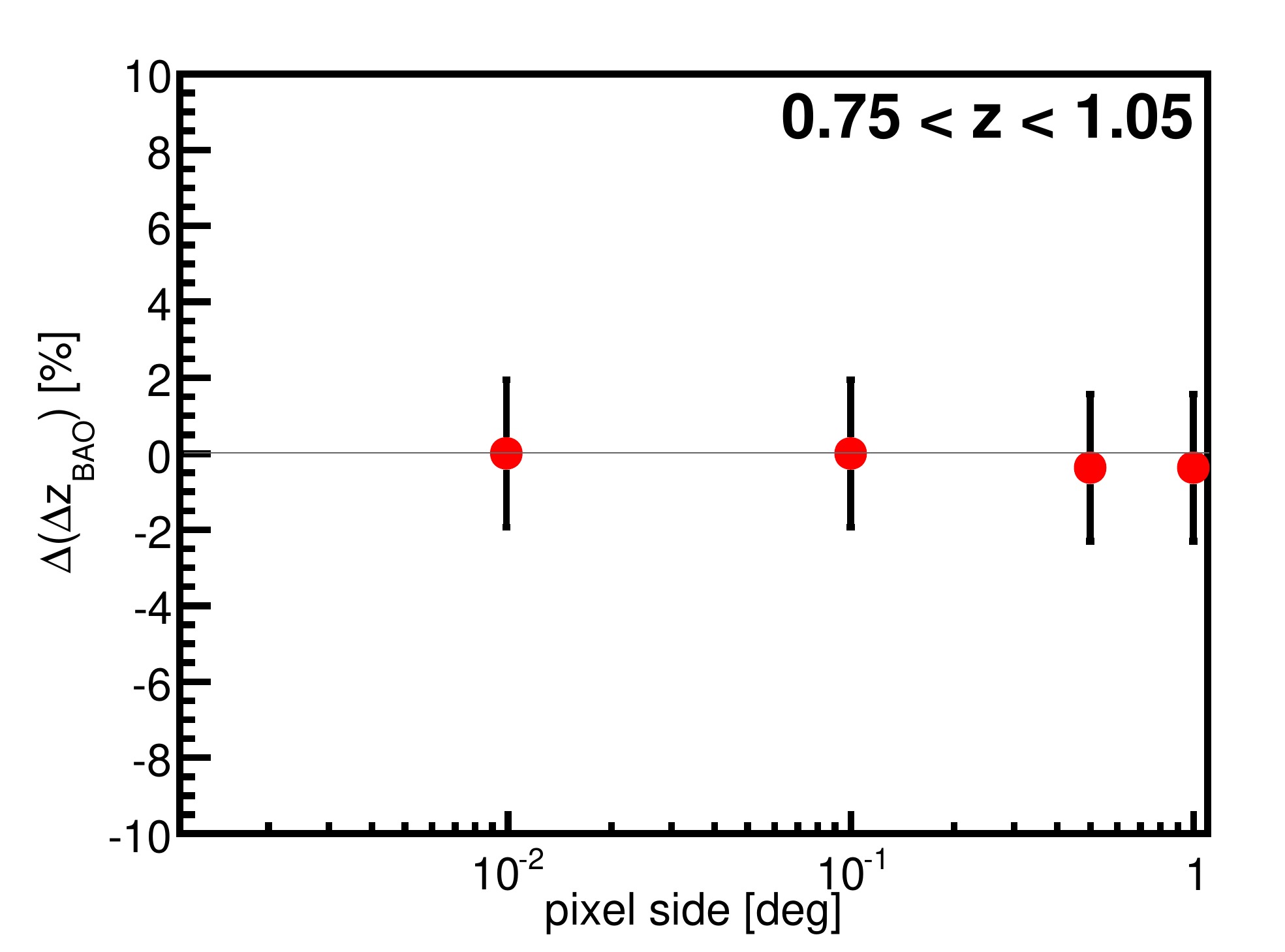} \\
    \includegraphics[width=0.40\textwidth]{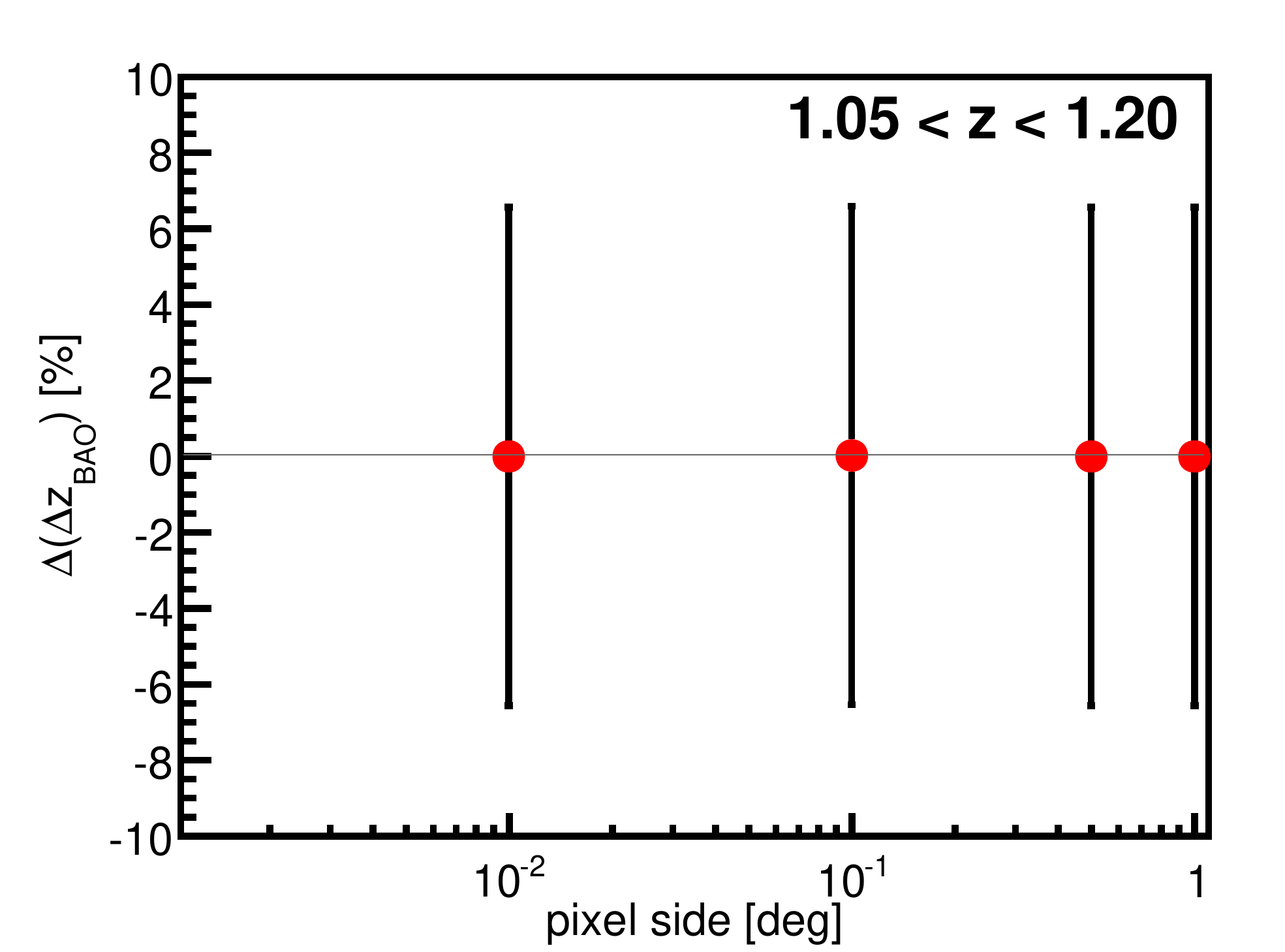} \\
  \end{tabular}
  \caption{Variation of the radial BAO scale determination as a function of the
           angular pixel size for different redshifts. Results are stable, and the maximum 
           variation is always of a few parts per mille, very well below 1\%, even if the 
           range in pixel sizes covers two orders of magnitude. The error bars indicate 
           the size of the statistical error for the nominal pixel size of 0.25 sq-deg, including 
           Poisson shot noise and cosmic variance, for the used simulation, that covers 1/8 of the sky. }
  \label{fig:syspixel}
\end{figure}

\subsubsection{Size of the Angular Pixel}

One specific systematic error associated with this method is the possible influence that the
size of the angular pixel has on the determination of the radial BAO scale, since it determines
which galaxies are considered collinear in the analysis. The effect of
different pixel sizes on the radial correlation function can be seen in 
Figure \ref{fig:pixelsize}. The radial correlation function clearly changes on small 
scales (Figure \ref{fig:pixelsize}, top), but this effect does not change the position of
the BAO peak (Figure \ref{fig:pixelsize}, bottom). The small scale effect comes from
the smoothing of the correlation function due to the inclusion in the calculation of galaxy 
pairs which are not exactly collinear. For a larger angular pixel, the effect is 
larger. However, the scale where this effect acts is fixed by the angular pixel size, which 
is very far from the BAO scale, that remains, therefore, unaffected, since the parametrization 
is able to absorb the change in the slope of the function.

In order to quantify this influence on the determination of the BAO scale as a 
systematic error, we have repeated the full analysis for different pixel sizes. The
obtained results are shown in Figure~\ref{fig:syspixel}. The radial BAO scale is recovered
with high precision for any angular pixel size, even for sizes as large as 1 degree, which
corresponds to a range of two orders of magnitude. The associated systematic error can be 
estimated to be $\delta(\Delta z_{BAO}) = 0.20 \%$, much smaller than the statistical error
for the nominal pixel size of 0.25 sq-deg, which is shown as error bars.

\subsubsection{Non-linearities}

Also the error due to the uncertainty in the goodness of the description provided
by the parametrization for different theoretical effects (non-linearities at 
the scale of the BAO peak) has been computed obtaining a global error 
of 0.10\%. This was estimated in a conservative way as the difference between the 
$\Delta z_{BAO}$ measured using linear and non-linear $\xi_{\parallel}(\Delta z)$, for
the same redshift bins of the analysis. Non-linearities are computed using the RPT
formalism \citep{2006PhRvD..73f3519C}, excluding mode-mode coupling, since it only 
affects small scales, far enough from the BAO scale. The contribution of these 
uncertainties to the systematic error can be estimated as 
$\delta(\Delta z_{BAO}) = 0.10 \%$.

\begin{figure}
  \centering
  \begin{tabular}{c}
    \includegraphics[width=0.40\textwidth]{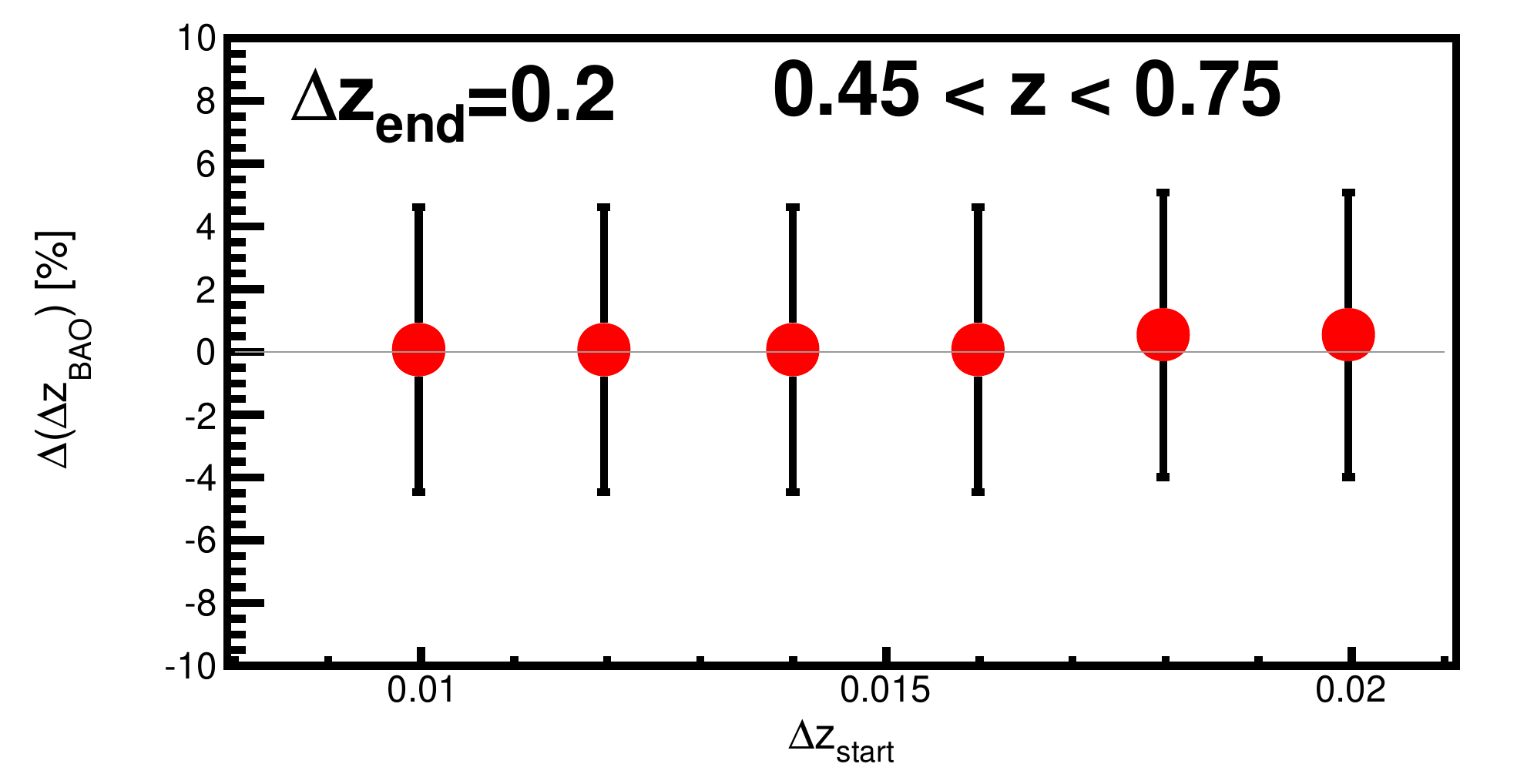}    \\
    \includegraphics[width=0.40\textwidth]{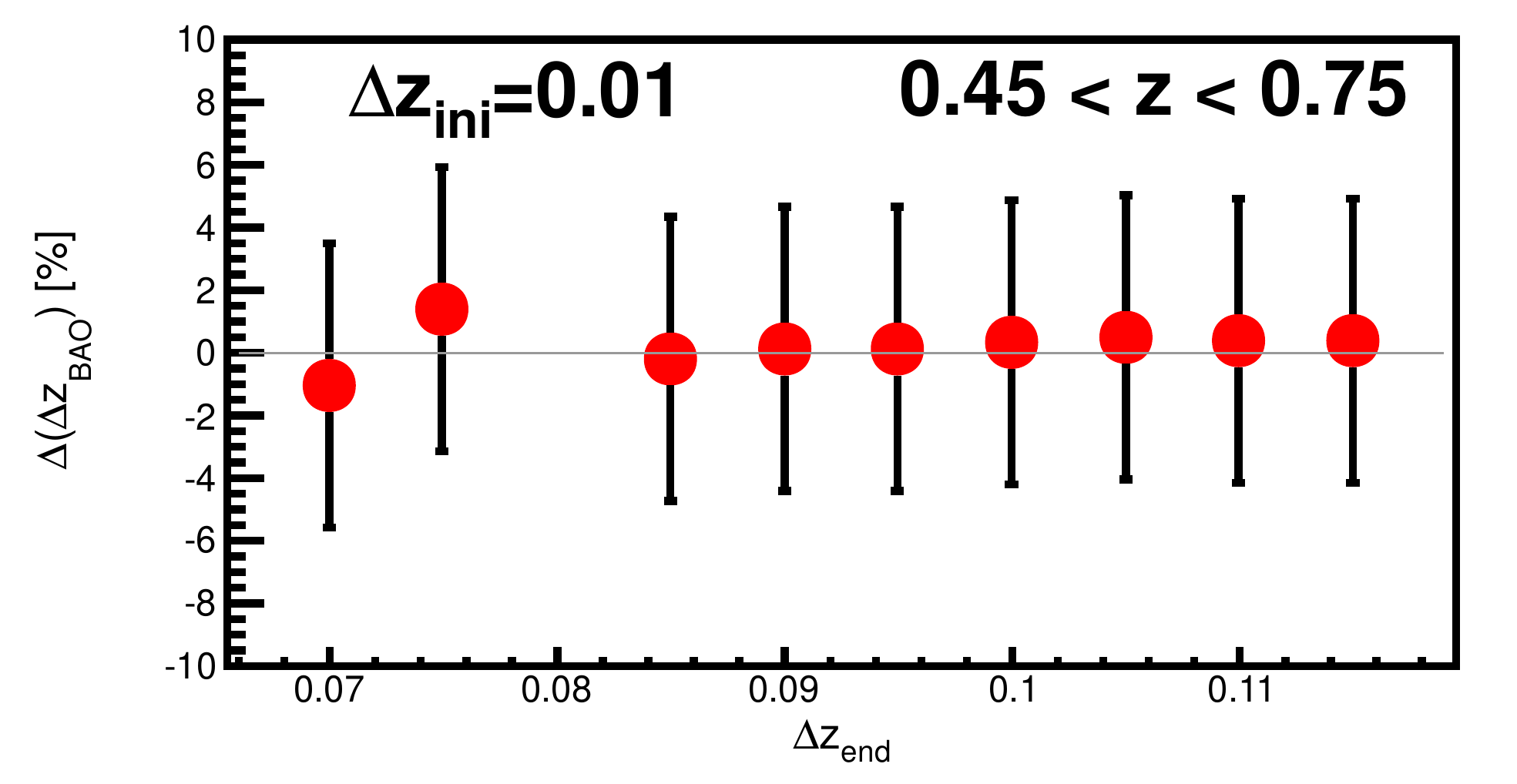}   \\
    \includegraphics[width=0.40\textwidth]{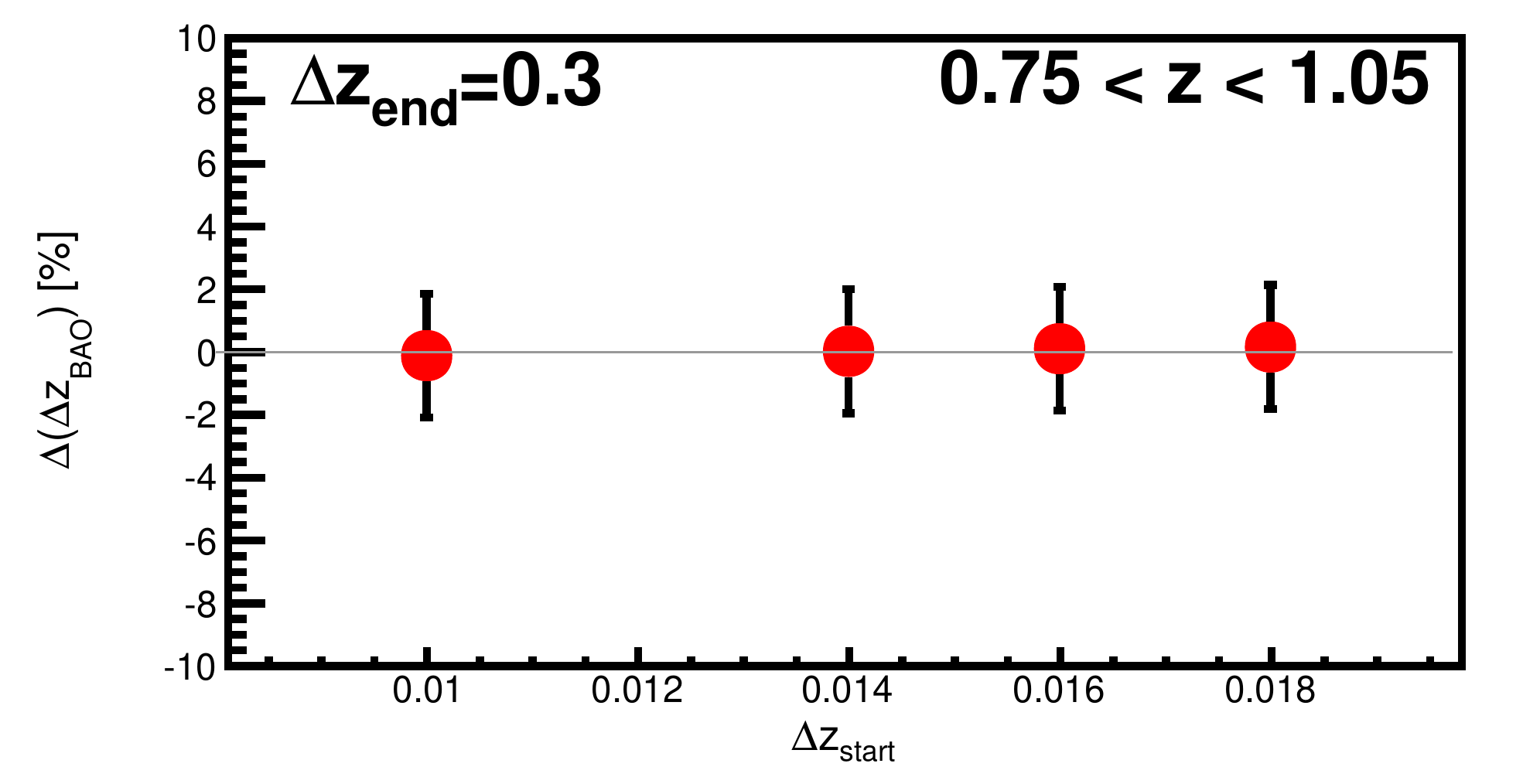}  \\
    \includegraphics[width=0.40\textwidth]{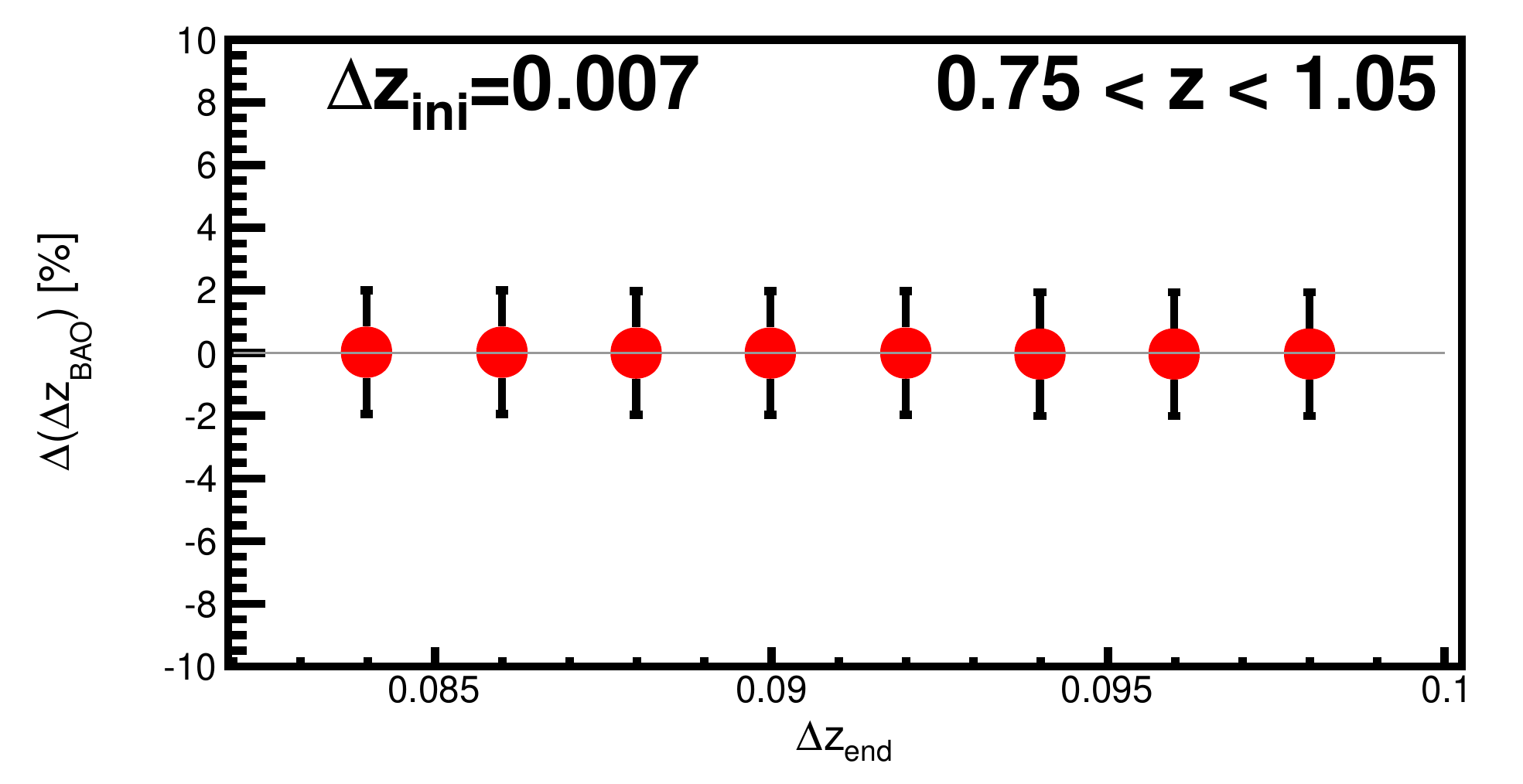} \\
    \includegraphics[width=0.40\textwidth]{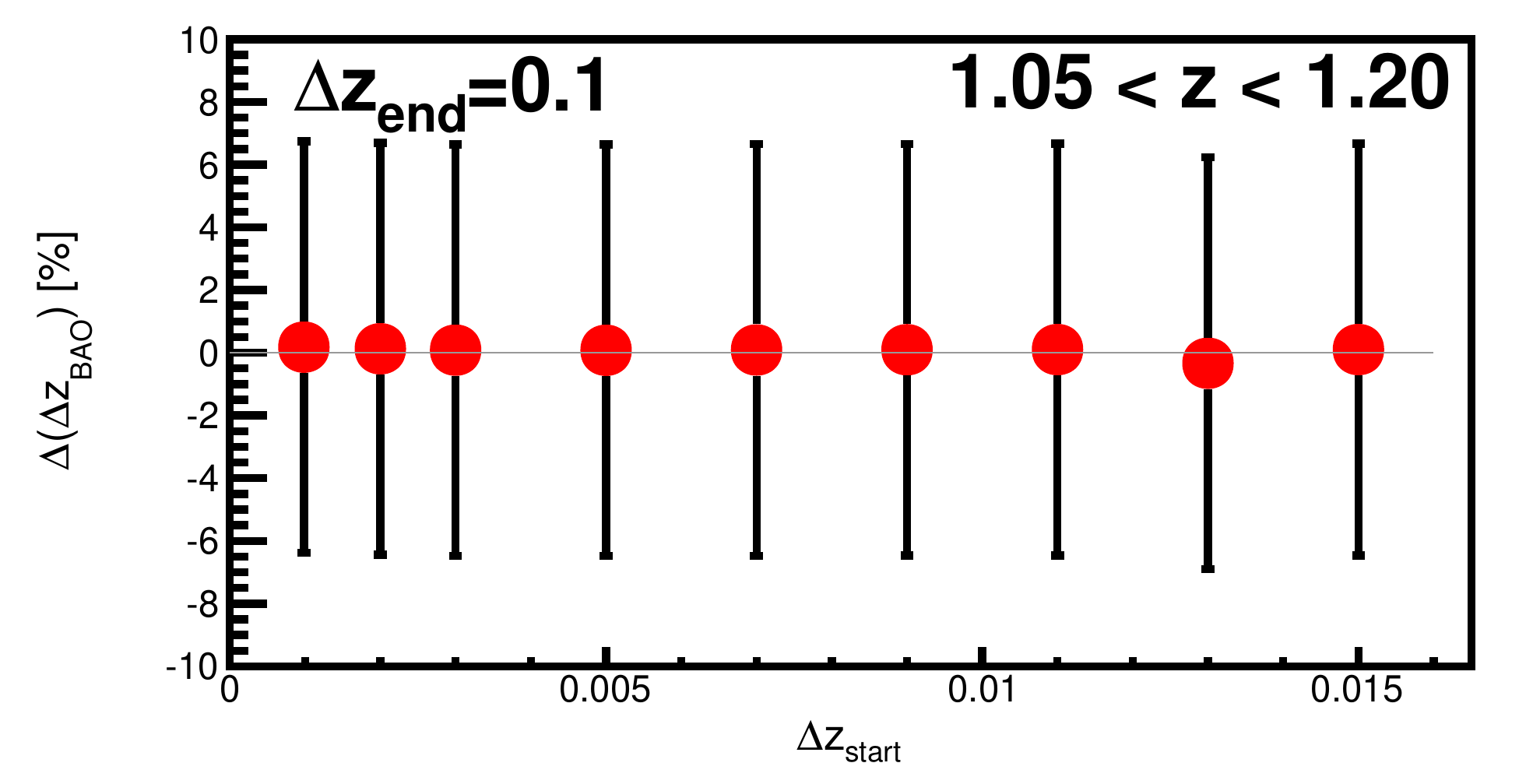}  \\
    \includegraphics[width=0.40\textwidth]{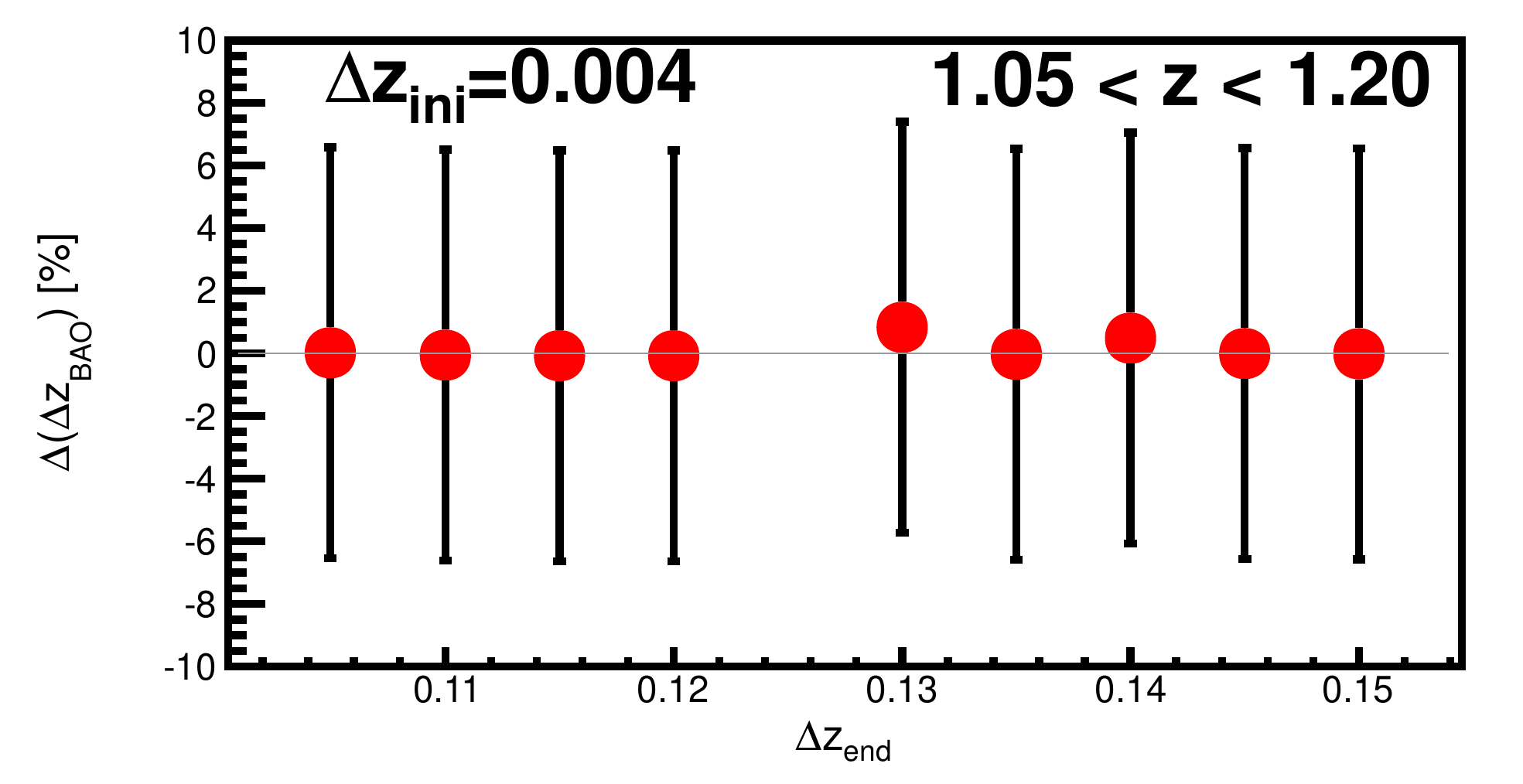} \\
  \end{tabular}
  \caption{$\Delta z_{BAO}$ evolution as a function of the starting and end point of the fitted 
           region. Results are stable, confirming that the systematic error is small. The error 
           bars indicate the size of the statistical error, including Poisson shot noise 
           and cosmic variance, for the used simulation, that covers 1/8 of the sky.}
  \label{fig:sysreg}
\end{figure}

\subsubsection{Galaxy Bias}

At the BAO scales it is a good approximation to consider that the bias 
is scale independent~\citep{2009MNRAS.392..682C,2011MNRAS.414..329C}. It affects the radial
correlation function not only as an overall normalization, but also through its effect 
in the redshift space distortions. Bias can influence the determination of the BAO scale
only through the changes in the goodness of the parametrization of the correlation 
function for different biases. In order to estimate the contribution of the galaxy
bias to the measurement of the radial BAO scale, we have repeated the analysis
with different values of the bias, to obtain the propagation of the uncertainty 
in the galaxy bias for the selected galaxy population to the measured value of
the radial BAO scale. The influence on the peak position is small, and we can
estimate the associated systematic error as $\delta(\Delta z_{BAO}) = 0.15 \%$.

Moreover, we have tested the effect of a scale dependent bias, introducing artificially
the effect in the correlation functions, using an approximate Q-model
with the determination of parameters from~\citet{2009MNRAS.392..682C}. The 
bias variation with $\Delta z$ in the fitted region of $\xi_{\parallel}(\Delta z)$ ranges 
from 1\% to 6\%, but the measurement of the BAO scale is insensitive to these changes. We
estimate the systematic error in the presence of a scale dependent bias 
as $\delta(\Delta z_{BAO}) = 0.20 \%$.

\subsubsection{Starting and End Point of the Fit}

To compute the systematic error associated to the parametrization method, we 
have done some further analysis on theoretical radial correlation functions with 
the same bin widths and central redshifts as those used in the analysis
of the MICE simulation. The error associated to the method comes from 
the possible influence in the obtained $\Delta z_{BAO}$ of the range of $\Delta z$ used 
to perform the fit. To evaluate the error, we have varied this range for the
3 redshift bins where we have a significant detection of the BAO scale, and performed the 
fit for each range. 

In the decision of the range to be fitted, we have to choose a starting point at angles 
smaller than the BAO peak, where the physics is determined by non-linearities, and 
an end point after the peak, beyond which the effects of cosmic variance may be 
relevant. By varying these two points we can study how much the result varies 
with this decision. Results can be seen in Fig.~\ref{fig:sysreg}, where
the obtained $\Delta z_{BAO}$ is shown for different starting points and 
end points  of the fit, for the 3 redshift bins. In all cases, the 
uncertainty is of the order of 0.1\%, which we assign as the associated systematic 
error. This uncertainty is much smaller than the statistical error, including Poisson 
shot noise and cosmic variance, which is depicted as error bars.

\subsubsection{Total Systematic Error}

The different sources of the systematic errors are completely independent, and therefore, we 
can compute the total systematic error by summing quadratically these contributions, resulting 
on a value of $\delta_{SYS}(\Delta z_{BAO}) = 0.33 \%$. 

There are some other potential systematic errors, the gravitational lensing 
magnification, which introduces a small correlation between redshift bins, or 
those mainly associated to the instrumental effects which could affect the used 
galaxy sample. However, these effects are expected to be very small and we have 
neglected them in this analysis.

\subsection{Cosmological Constraints}
\label{sec:cosmopars}

The evolution of the measured radial BAO scale, including the systematic errors, with 
redshift is shown in Figure~\ref{fig:dzBAOvsz}. The cosmological model of the simulation 
is the solid line. The recovered BAO scale is perfectly compatible with the true 
model, demonstrating that the method works.

\begin{figure}
  \includegraphics[width=0.49\textwidth]{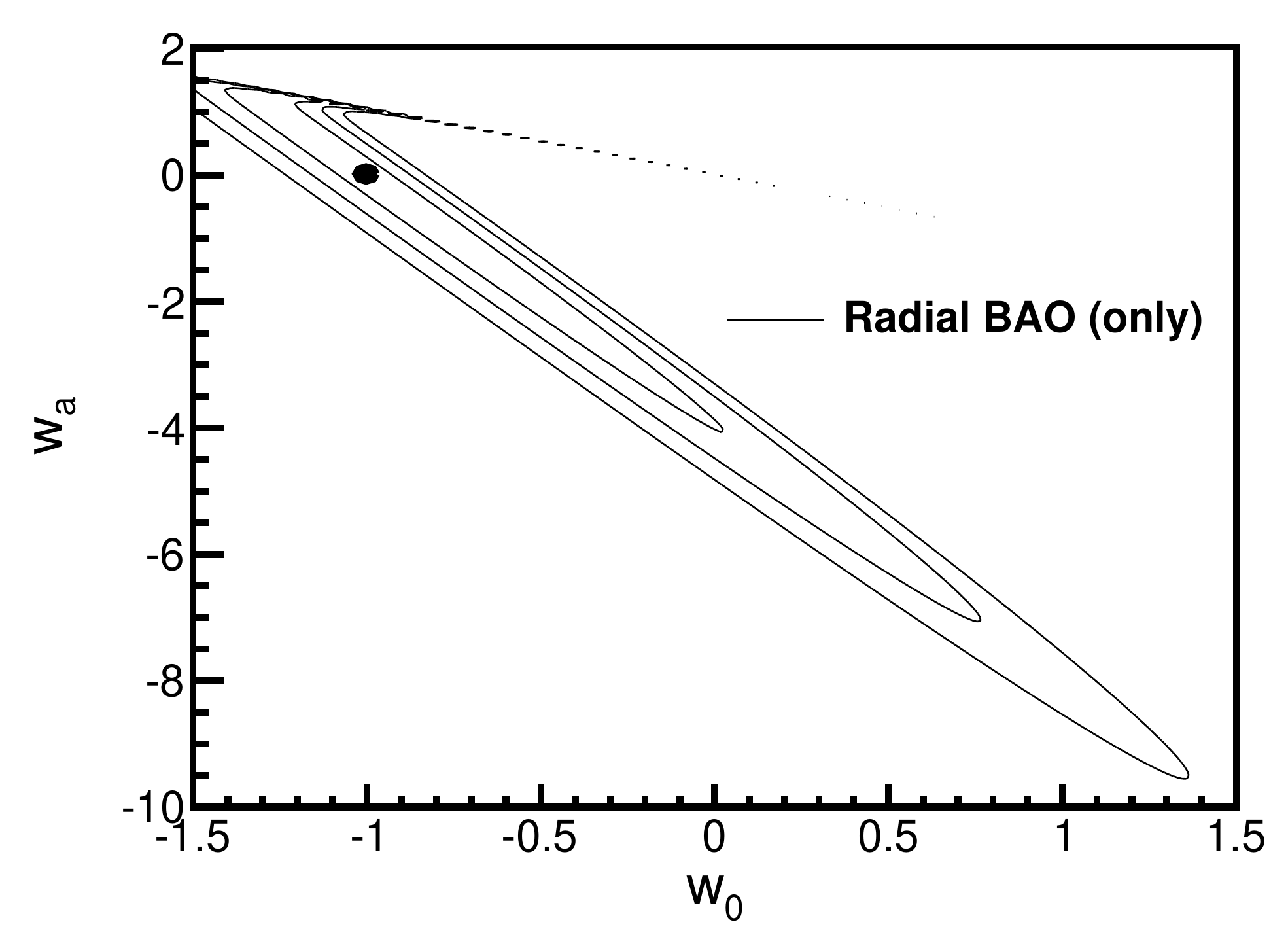}
  \includegraphics[width=0.49\textwidth]{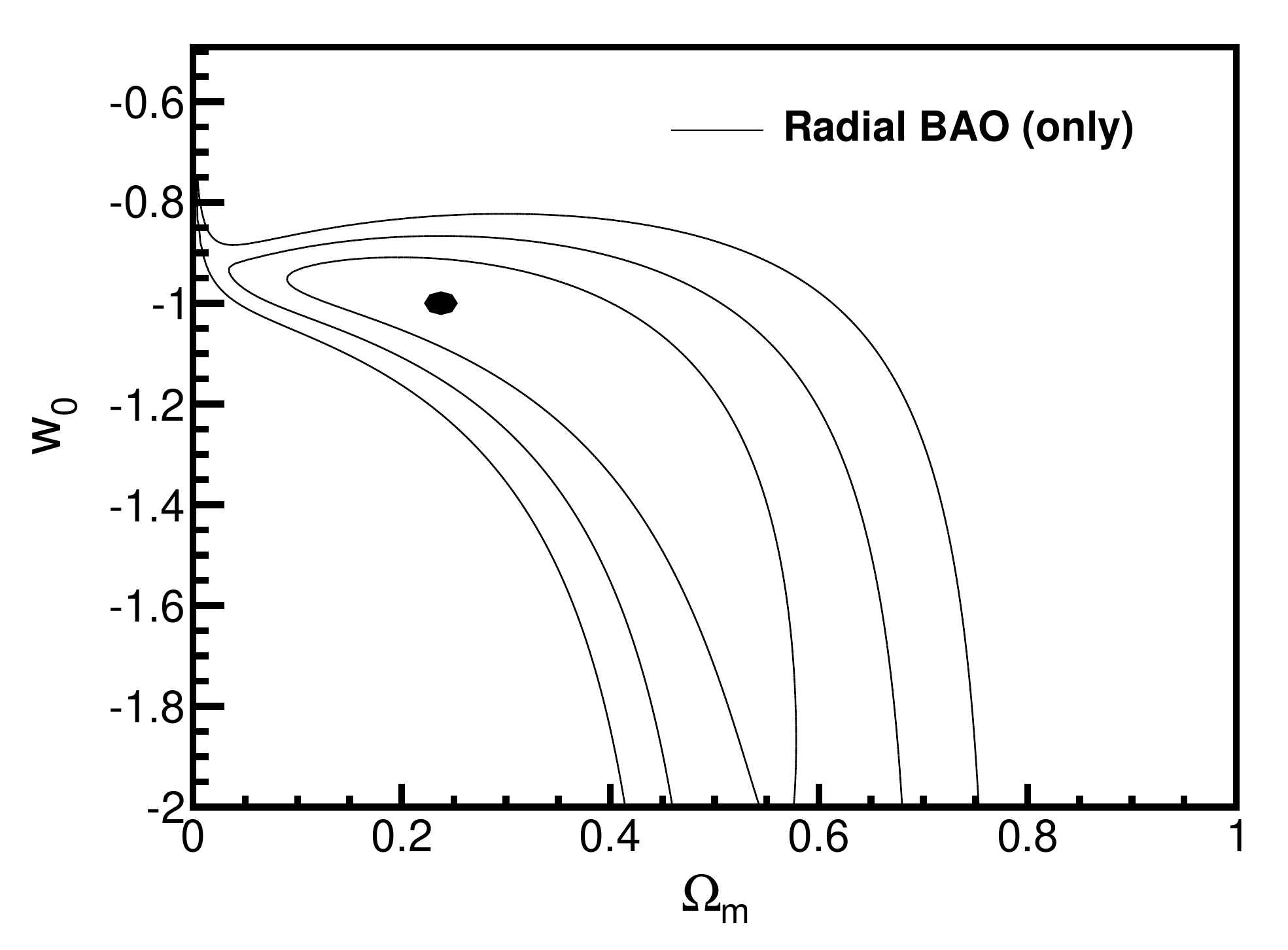}
  \caption{Contours at $1\sigma$, $2\sigma$ and $3\sigma$ C. L. on the plane $(w_0, w_a)$ (top)
           and on the plane $(\Omega_M, w_0)$ (bottom) obtained from the analysis of the radial 
           BAO scale. The dot shows the value of the parameters for the MICE cosmology. No 
           combination with any other cosmological probe is included. The other parameters 
           have been fixed to the values of the MICE cosmology.}
  \label{fig:cosmoparsfromradialBAO}
\end{figure}

When these measurements are translated into constraints on the cosmological parameters, 
we obtain the results depicted in Figure~\ref{fig:cosmoparsfromradialBAO}, where the contours
for $1\sigma$, $2\sigma$ and $3\sigma$ C. L. in the $(w_0, w_a)$ plane are shown at the top 
panel and the bottom panel shows the same contours in the $(\Omega_M, w_0)$ plane. To obtain 
the constraints on the cosmological parameters, we have performed a $\chi^2$ fit to the
evolution of the measured radial BAO scale with the redshift to the model, where

\begin{eqnarray}
\Delta z_{BAO} = r_{S}(\Omega_{M}, w_0, w_a...)~H(z,\Omega_{M}, w_0, w_a...),
\end{eqnarray}

\noindent
where $r_{S}$ is the sound horizon scale at the baryon drag epoch and 
$H(z,\Omega_{M}, w_0, w_a...)$ is the Hubble parameter. We leave free those 
cosmological parameters which are shown in the figures, while all
other parameters have been kept fixed to their values for the simulation. The cosmology 
of the simulation is recovered, and the plot shows the sensitivity of the radial BAO 
scale alone, since no other cosmological probe is included in these constraints.

\begin{figure}
  \includegraphics[width=0.49\textwidth]{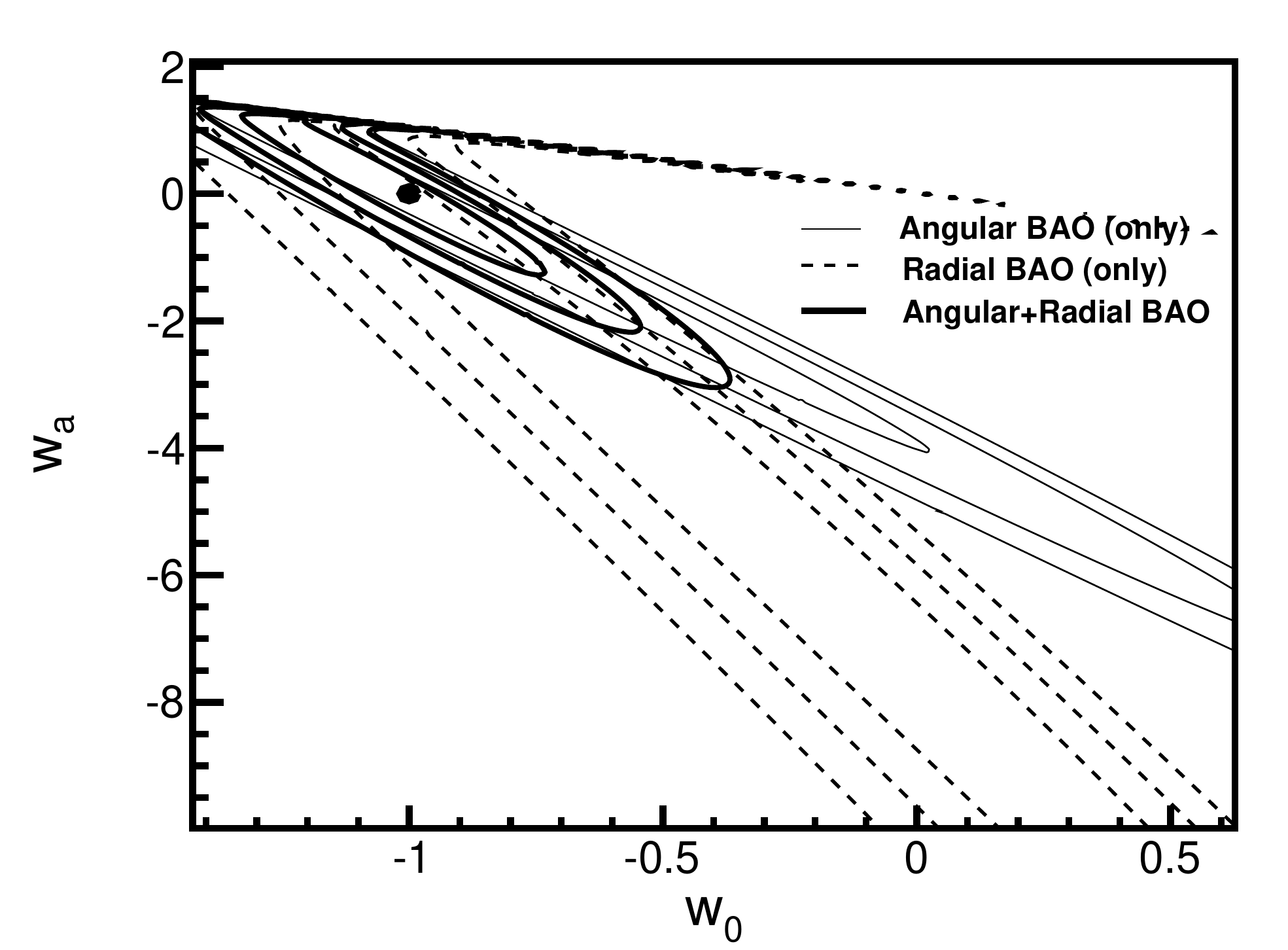}
  \includegraphics[width=0.49\textwidth]{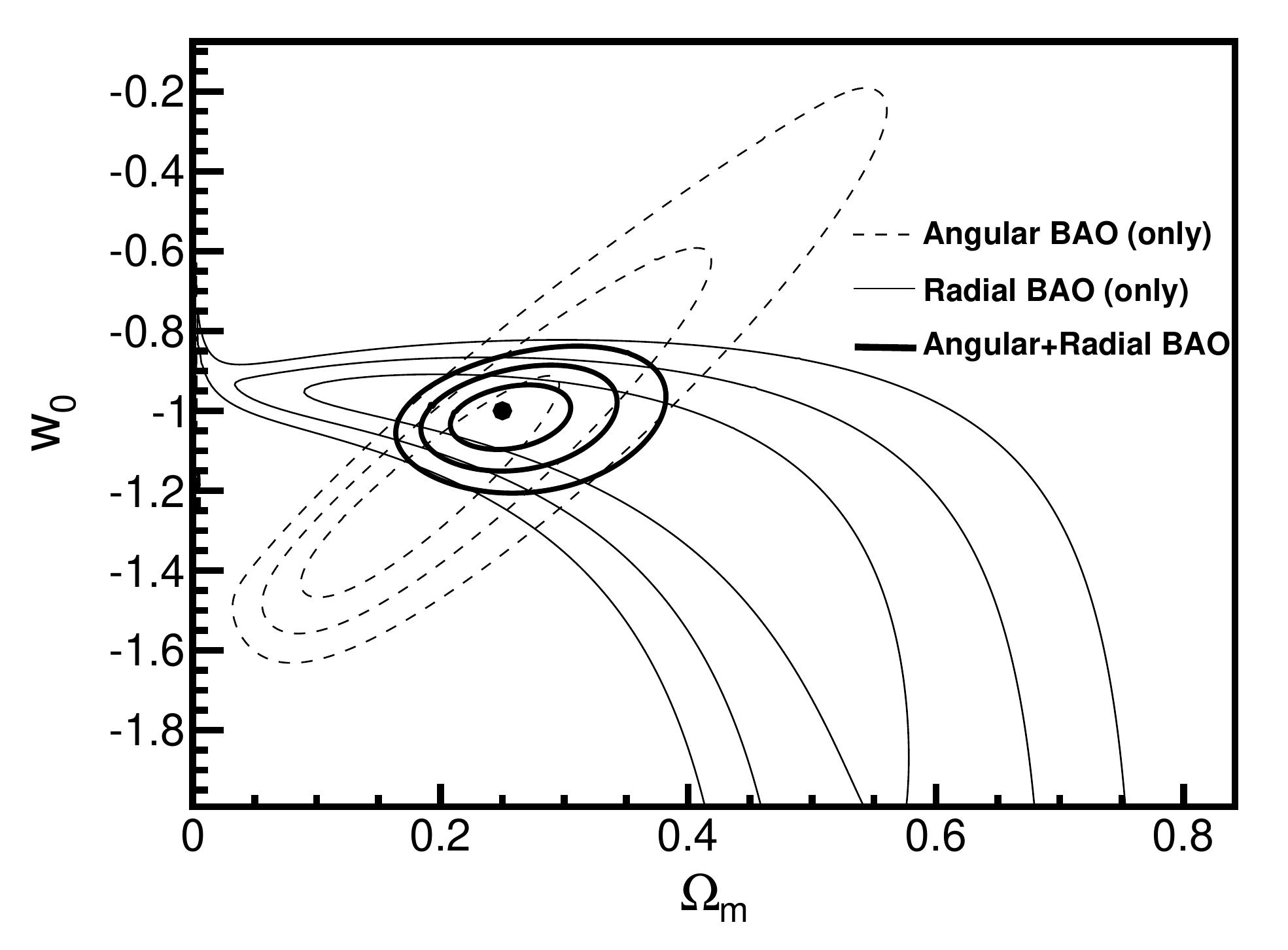}
  \caption{Contours at $1\sigma$, $2\sigma$ and $3\sigma$ C. L. on the plane $(w_0, w_a)$ (top)
           and on the plane $(\Omega_M, w_0)$ (bottom) from 
           radial BAO (thin solid lines), angular BAO (thin dashed lines) and the combination 
           of both (thick solid lines). The dot shows the value of the parameters for the MICE 
           cosmology. No other cosmological probe is included in this 
           result, showing the high sensitivity that the BAO standard ruler can achieve. The 
           other parameters have been fixed to their values in the MICE cosmology.}
  \label{fig:cosmoparstotalBAO}
\end{figure}

\subsection{Combination of Radial and Angular BAO Scales}
\label{sec:fullbao}

We have applied the method of \cite{2011MNRAS.411..277S} to determine the angular BAO 
scale for the same simulation. The results of this analysis are presented in the 
Appendix~\ref{sec:Angular_BAO}. We have combined the results of this analysis with the 
radial BAO scale, using the same approach of the previous section. For the angular 
analysis, the BAO scale is described as

\begin{eqnarray}
\theta_{BAO} = \frac{r_{S}(\Omega_{M}, w_0, w_a...)}{(1+z)~d_{A}(z,\Omega_{M}, w_0, w_a...)},
\end{eqnarray}

\noindent
where $d_{A}(z,\Omega_{M}, w_0, w_a...)$ is the angular diameter distance. The constraints 
on the $(w_0, w_a)$ plane coming from this determination of the angular BAO scale can be seen in 
Figure~\ref{fig:cosmoparstotalBAO} (top) as thin dashed lines, and correspond to the 
$1\sigma$, $2\sigma$ and $3\sigma$ C. L. contours. The 
sensitivity of the angular BAO scale is complementary to that of the radial BAO, shown 
as the thin solid lines. When combined, the contours represented by the thick solid lines 
are found. The same constraints for the $(\Omega_M, w_0)$ plane are shown in the bottom 
panel of Figure~\ref{fig:cosmoparstotalBAO}. These constraints are as 
precise as what is usually quoted for the BAO standard ruler, which is based on the use 
of the monopole of the 3-D correlation function. It is important to remark that these 
constraints are obtained ONLY with the BAO standard ruler, independently of any other 
cosmological probe, which shows the real power of the standard ruler method when the 
full information is used.

\begin{figure}
  \includegraphics[width=0.49\textwidth]{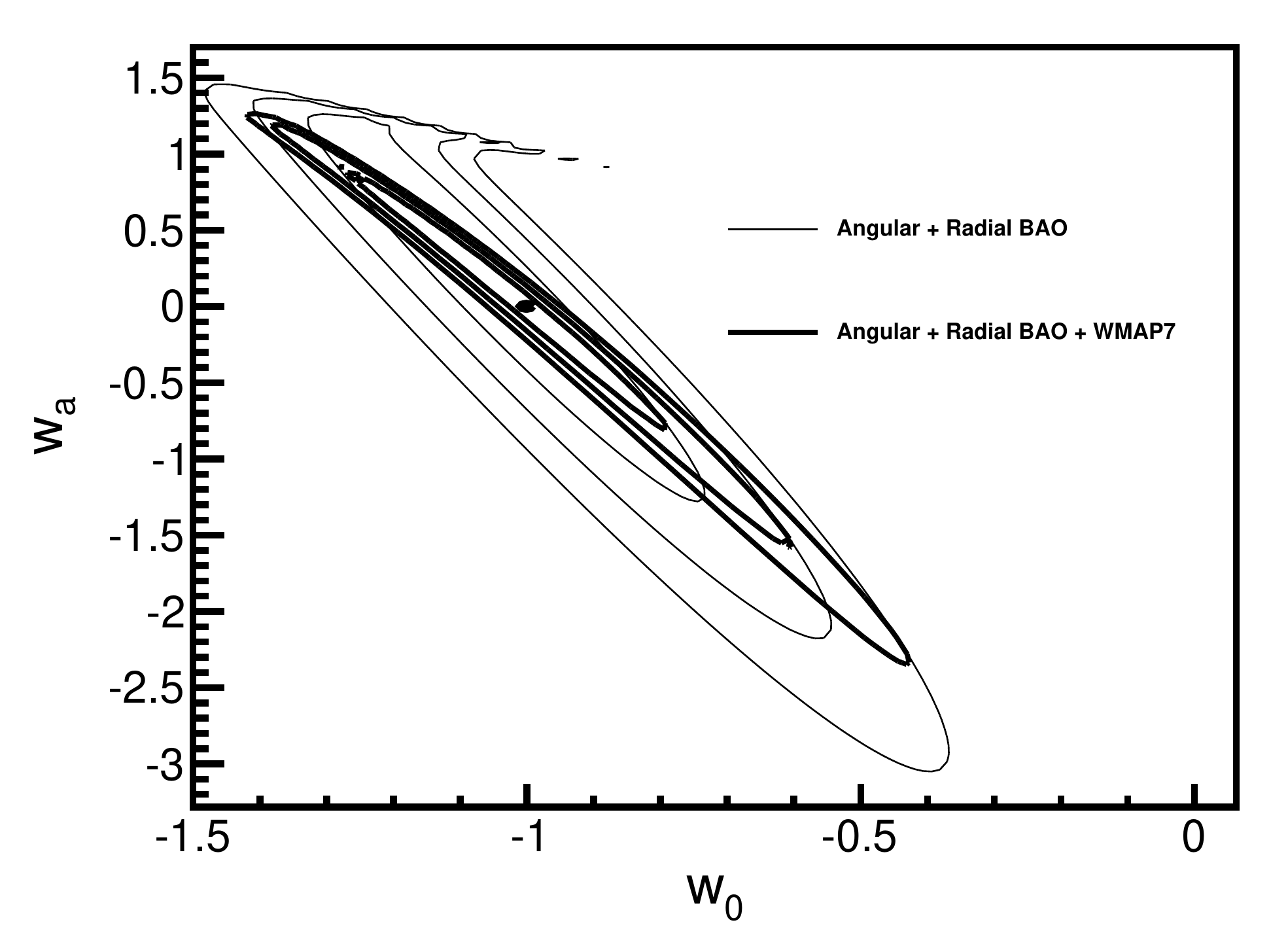}
  \includegraphics[width=0.49\textwidth]{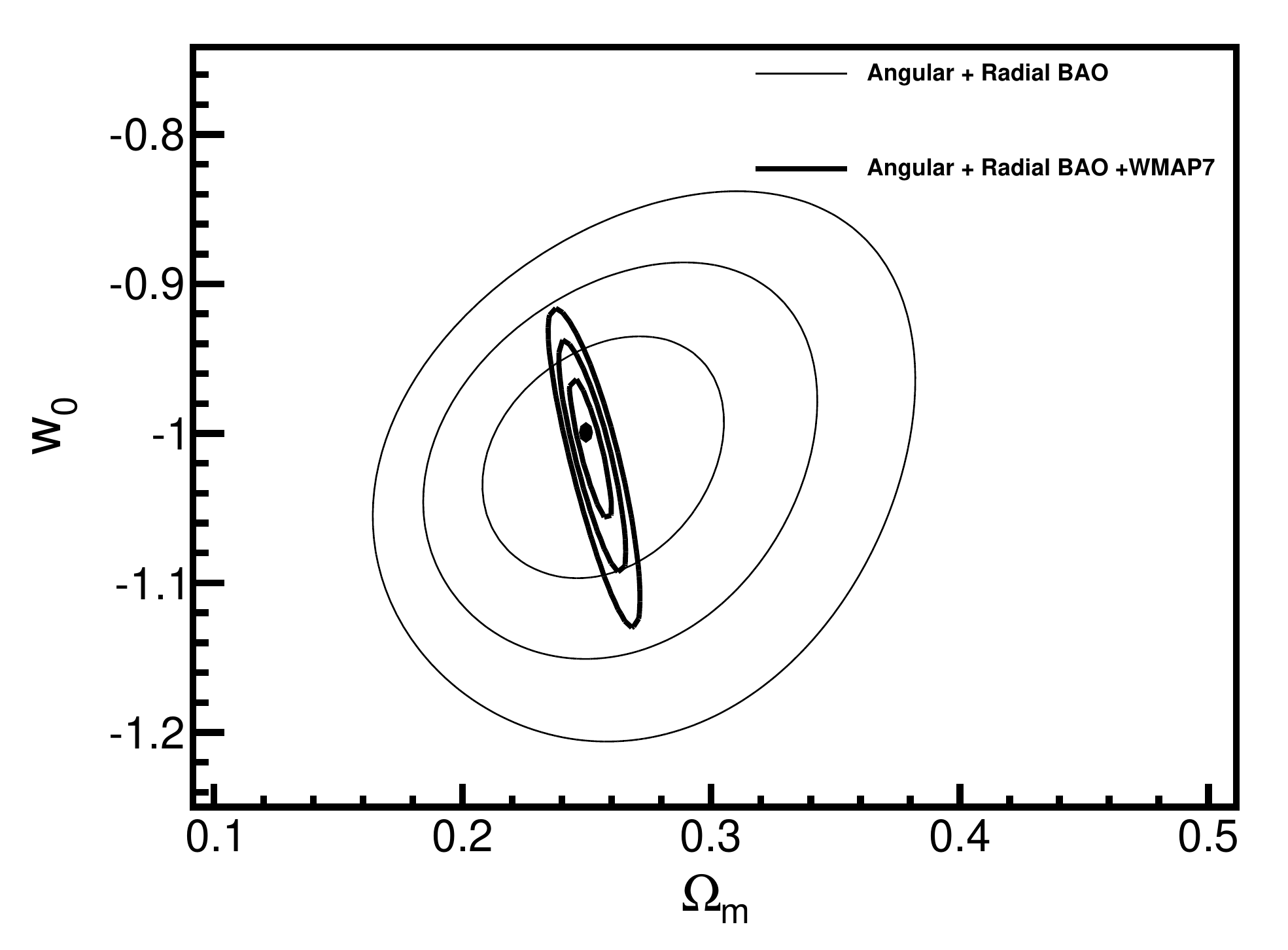}
  \caption{Contours at $1\sigma$, $2\sigma$ and $3\sigma$ C. L. on the plane 
           $(w_0, w_a)$ (top) and on the plane $(\Omega_M, w_0)$ (bottom) from  
           BAO (both radial and angular), depicted as thin solid lines, and adding 
           also the CMB measurements, depicted as thick solid lines. The dot shows 
           the value of the parameters for the MICE cosmology. The 
           covariance matrix of WMAP7 has been used, while the central value of 
           the measurement has been taken at the MICE cosmology. The other parameters 
           have been fixed to their values in the MICE cosmology.}
  \label{fig:wOmBAOCMB}
\end{figure}

We provide also the combined result of BAO, both radial and angular, with
the distance measurements from CMB using the WMAP7 covariance 
matrix \citep{2010arXiv1001.4538K} and assuming the measurements correspond to the
MICE cosmology. The combination has been performed following the procedure
as detailed in \cite{2009ApJS..180..330K}. The corresponding contours at 
$1\sigma$, $2\sigma$ and $3\sigma$ C. L. are presented in 
Figure~\ref{fig:wOmBAOCMB} both in the plane $(w_0, w_a)$ (top) and in the
plane $(\Omega_M, w_0)$ (bottom) as thick solid lines, and compared with the
result using only BAO (both radial and angular), which is presented as the 
thin lines. As before, the other parameters have been kept fixed. There is an 
important improvement in the precision of the determination of the corresponding
parameters in both cases, larger in the $(\Omega_M, w_0)$ plane, showing
that a precise measurement can be achieved when all the information provided by
the BAO scale is included in the fit.

\subsection{Comparison with Other Methods}
\label{sec:comparison}

In order to compare our results, obtained combining radial and angular BAO 
scale determinations, with the standard approach of measuring the position of the sound 
horizon scale in the monopole of the two-point correlation function, we performed on our 
mock catalogue the same analysis that was carried out to obtain the latest BAO detection 
by the BOSS collaboration \citep{2012MNRAS.427.3435A}. For this we calculated the 
three-dimensional correlation function $\xi(r)$ using the Landy and Szalay
estimator in the three wide redshift bins used for the analysis of the radial BAO
($0.45<z<0.75$, $0.75<z<1.10$ and $1.10<z$). We chose these wide bins in order to maximize
the number of pairs that contribute to the measurement of the monopole, since this is one of
the key advantages of the standard method. In order to calculate $\xi(r)$, redshifts must
be translated into distances. We have used the true cosmology of the MICE simulation to 
ensure that our results will not be biased by this choice.
    
The covariance matrix for the monopole was calculated using the Gaussian approach described
by \citet{2013MNRAS.431.2834X}. This calculation was, again, validated
by comparing it with the errors computed from subsamples of the total catalogue, and both
estimations were found to be compatible within the range of scales needed for the analysis.

\begin{figure}
  \centering
  \includegraphics[width=0.49\textwidth]{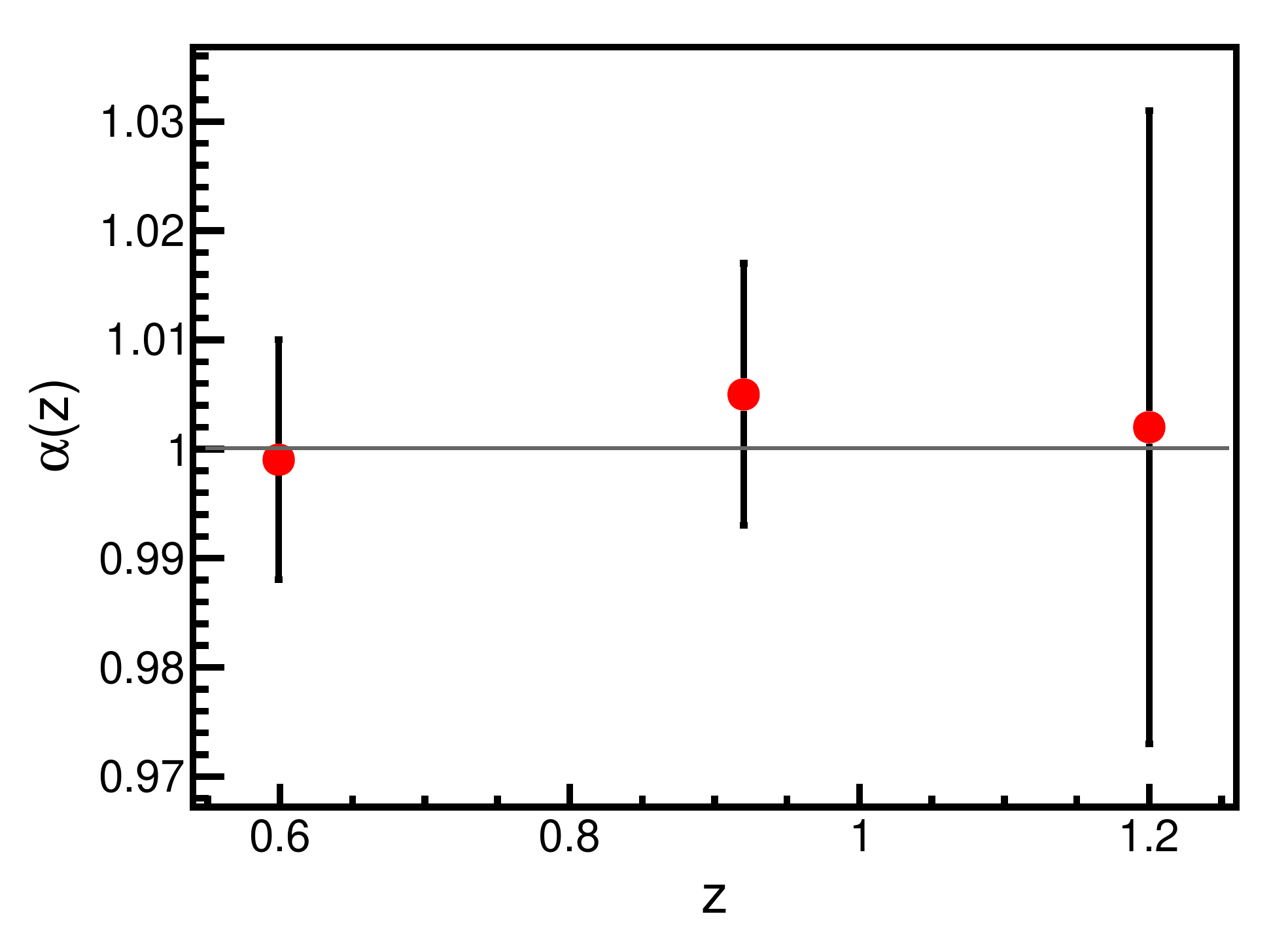}
  \caption{Values of the scaling parameter $\alpha$ measured from the three bins of the MICE mock
           catalogue. The input cosmology ($\alpha\equiv1$) is recovered well within errors.}
  \label{fig:alpha_monos}
\end{figure}

As is done in \citep{2012MNRAS.427.3435A}, we fit the model
\begin{equation}\label{eq:fit_modl}
  \xi_{\rm fit}(r)=B^2\,\xi_{\rm th}(\alpha\,r)+a_0+\frac{a_1}{r}+\frac{a_2}{r^2},
\end{equation}

\noindent
to the estimated correlation monopoles. Here $\xi_{\rm th}(r)$ is a template theoretical
correlation function corresponding to the fiducial cosmological model used to translate redshifts
into distances in the survey (in our case the MICE cosmology). This template was calculated from
the {\tt CAMB} linear power spectrum for the MICE cosmology, and corrected for non-linearities
via the RPT damping factor. We are mainly interested in the fitting parameter $\alpha$, which
relates real and fiducial scales:

\begin{equation}
  \frac{d_V(z)}{r_s}=\alpha\,\frac{d_V^{\rm fid}(z)}{r_s^{\rm fid}},
\end{equation}

\noindent
where $d_V(z)\equiv((1+z)^2\,d_A^2(z)\,z/H(z))^{1/3}$ is the volume-averaged distance defined
by \cite{2005ApJ...633..560E}. Since the true cosmology was used to translate redshifts into
distances, the value of $\alpha$ must be compatible with 1 (within errors). The statistical
uncertainty in $\alpha$ was calculated following the same method used in
\cite{2012MNRAS.427.3435A}. 
 
We have not studied the different sources of systematic errors for this measurement, and no
systematic contribution has been added to the errors. On the one hand this provides a more
conservative comparison with our approach, since the results quoted in section \ref{sec:fullbao}
do contain systematics. On the other hand, there exist several potential systematics that are
specific for the standard method, such as the effect of the fiducial cosmology used to obtain
the three-dimensional positions of the galaxies, or the choice of template used to perform
the fit. Studying this effect would be extremely interesting, but we have postponed this
analysis for a future work. As we have seen before, the systematic errors that are common to
both approaches (bias, RSDs, non-linearities, fitting limits) are clearly subdominant
compared to the statistical uncertainties.

\begin{figure}
  \centering
  \includegraphics[width=0.49\textwidth]{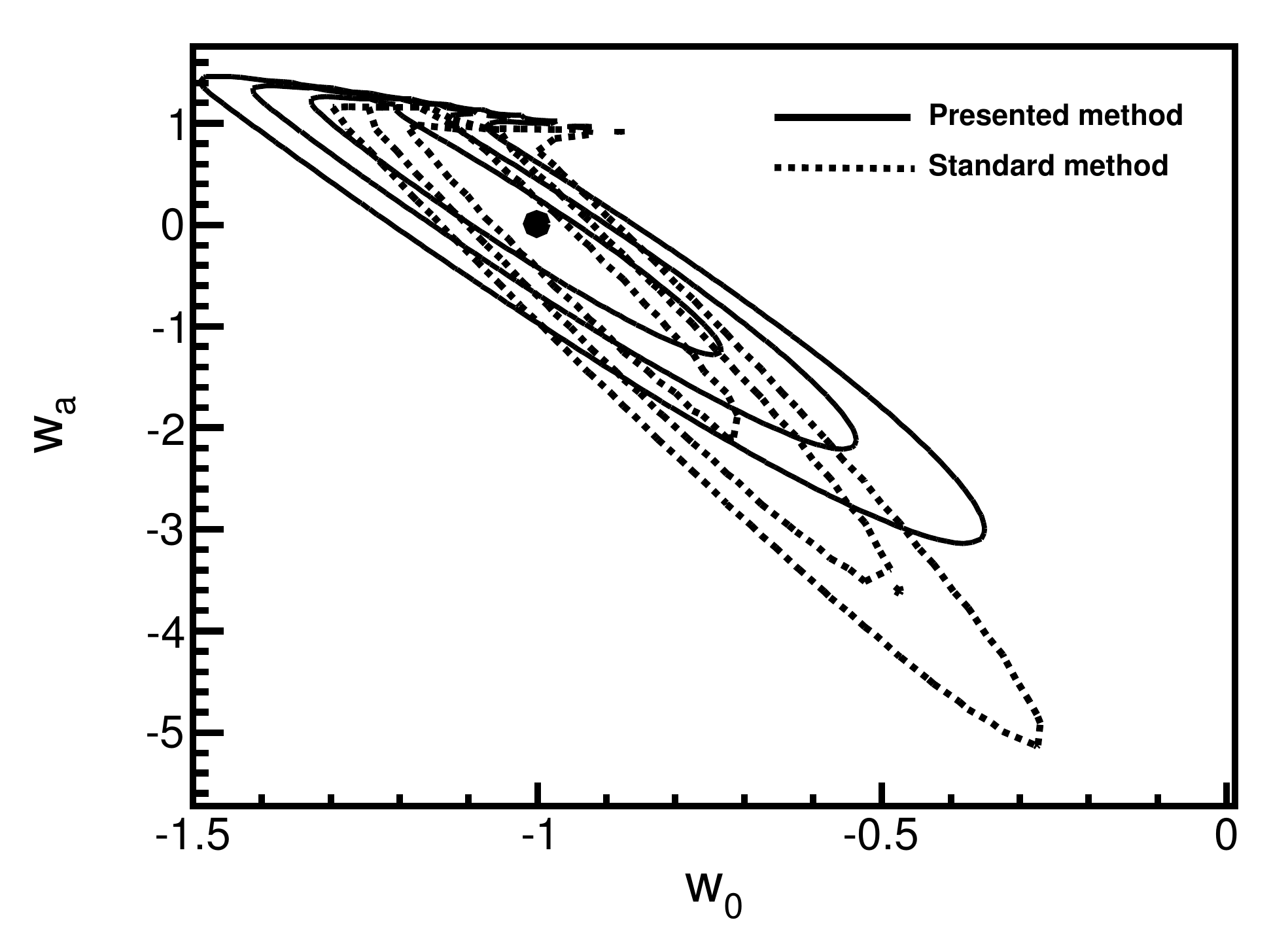}
  \includegraphics[width=0.49\textwidth]{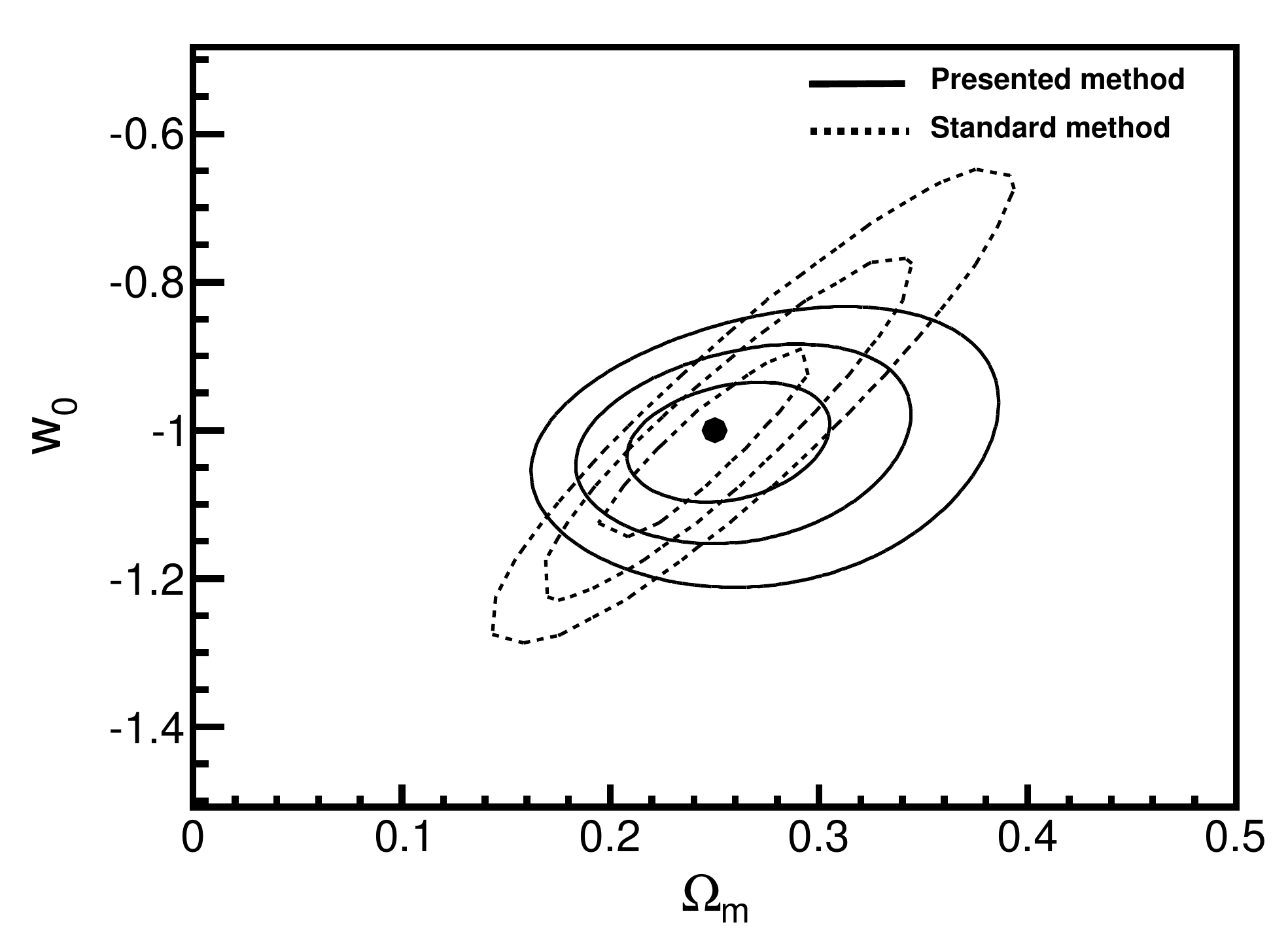}
  \caption{Contours at 1-$\sigma$, 2-$\sigma$ and 3-$\sigma$ confidence level in the planes
           $(w_0,w_a)$ (left panel) and $(\Omega_{\rm M},w_0)$ (right panel). The solid contours
           correspond to our combination of independent radial and angular BAO information, while
           the dashed lines correspond to the results drawn from the standard analysis of the
           angle-averaged BAO. The different correlation between parameters is due to the different
           treatment of the redshift space distortions.}
  \label{fig:contours_monos}
\end{figure}

The cosmological constraints drawn  in the $(\Omega_{\rm M},w_0)$ and
$(w_0,w_a)$ planes from the values of $\alpha$ measured from the correlation
functions are shown in Figure~\ref{fig:contours_monos}. The figure also shows the contours 
corresponding to the combination of radial and angular information, described in the previous 
section, for comparison. Plots show that the constraining power of both methods is 
very similar. There is a degenerate direction in the $(\Omega_{\rm M},w_0)$ plane for the 
standard method, which coincides with the orientation of the contours for the angular BAO shown in 
Figure~\ref{fig:cosmoparstotalBAO} (bottom panel). This is a reasonable result: most of the 
information in the angle-averaged BAO signature comes from the angular part, since there are 
two transverse dimensions and only one longitudinal. On the other hand, our combined approach 
seems to be able to obtain better constraints on the evolution of the dark energy equation of 
state. This could be due to the fact that the radial BAO enables us to measure the evolution 
of the expansion rate alone, which is a local quantity, unlike the angular diameter distance, which 
is an integrated one depending on the expansion history. Although the degenerate directions are
very similar for both methods, they are not exactly the same. The reason is the different
treatment of the redshift space distortions, which are ignored, to first approximation, in the
standard method when the angular average is performed. However, they are fully taken into account
in our proposal, where the angular and radial correlation functions have very different
shapes, mainly because of the redshift space distortions.

\section{Conclusions}
\label{sec:conclusions} 

We have developed a new method to measure the BAO scale in the radial two-point correlation 
function. This method is adapted to the observational characteristics of galaxy surveys, where
only the angular position on the sky and the redshift are measured for each galaxy. The sound 
horizon scale can be recovered from the non-linear radial correlation functions to a 
very high precision, only limited by the volume of the considered survey, since the systematic 
uncertainties associated to the determination of the BAO scale are very small, around 0.3\%. On 
the other hand, the method is fully cosmology independent, since it relies only on observable
quantities and, consequently, its results can be analyzed in any cosmological model.

The method has been tested with a mock catalogue built upon a large N-body simulation provided 
by the MICE collaboration, in the light cone and including redshift space distortions. The true 
cosmology is recovered within 1-$\sigma$. An evaluation of the main systematic errors has 
been included in this study, and we find that the method is very promising and very 
robust against systematic uncertainties. Note that this analysis over the MICE simulations is done 
on dark matter particles, instead of galaxies. We believe this simplification is not an 
essential limitation to the method presented here, as we have shown that both the modeling and 
the error analysis are quite generic. 

We have compared the cosmological constraints obtained by combining radial and angular BAO
information with those obtained by performing the standard analysis of the angle-averaged BAO
signature on the same dataset and with the same fitting technique. Both methods seem to yield 
comparable constraints, with the advantage that our method is entirely based on purely observable 
quantities (redshifts and angles) and is therefore completely model-independent.

\section*{Acknowledgements}
\label{sec:acknowledgements}  

We acknowledge useful comments from Enrique Gazta\~naga and Pablo Fosalba that helped to improve 
this work. We acknowledge the use of data from the MICE simulations, kindly provided by the MICE 
collaboration. We also thank the anonymous referee for the comments and suggestions. Funding 
support for this work was provided by  the Spanish Ministry of Science and Innovation (MICINN) 
through grants AYA2009-13936-C06-03, AYA2009-13936-C06-06 and through the Consolider 
Ingenio-2010 program, under project CSD2007-00060. DA acknowledges support from a JAE-predoc 
contract. JGB and DA acknowledge financial support from the Madrid Regional Government 
(CAM) under the program HEPHACOS S2009/ESP-1473-02. 


\appendix
\section{Gaussian approach to estimate the covariance matrix of the 3-D correlation function}
\label{sec:error3-D}

\begin{figure*}
  \centering
  \leavevmode
  \begin{tabular}{ccc}
    \includegraphics[width=0.30\textwidth]{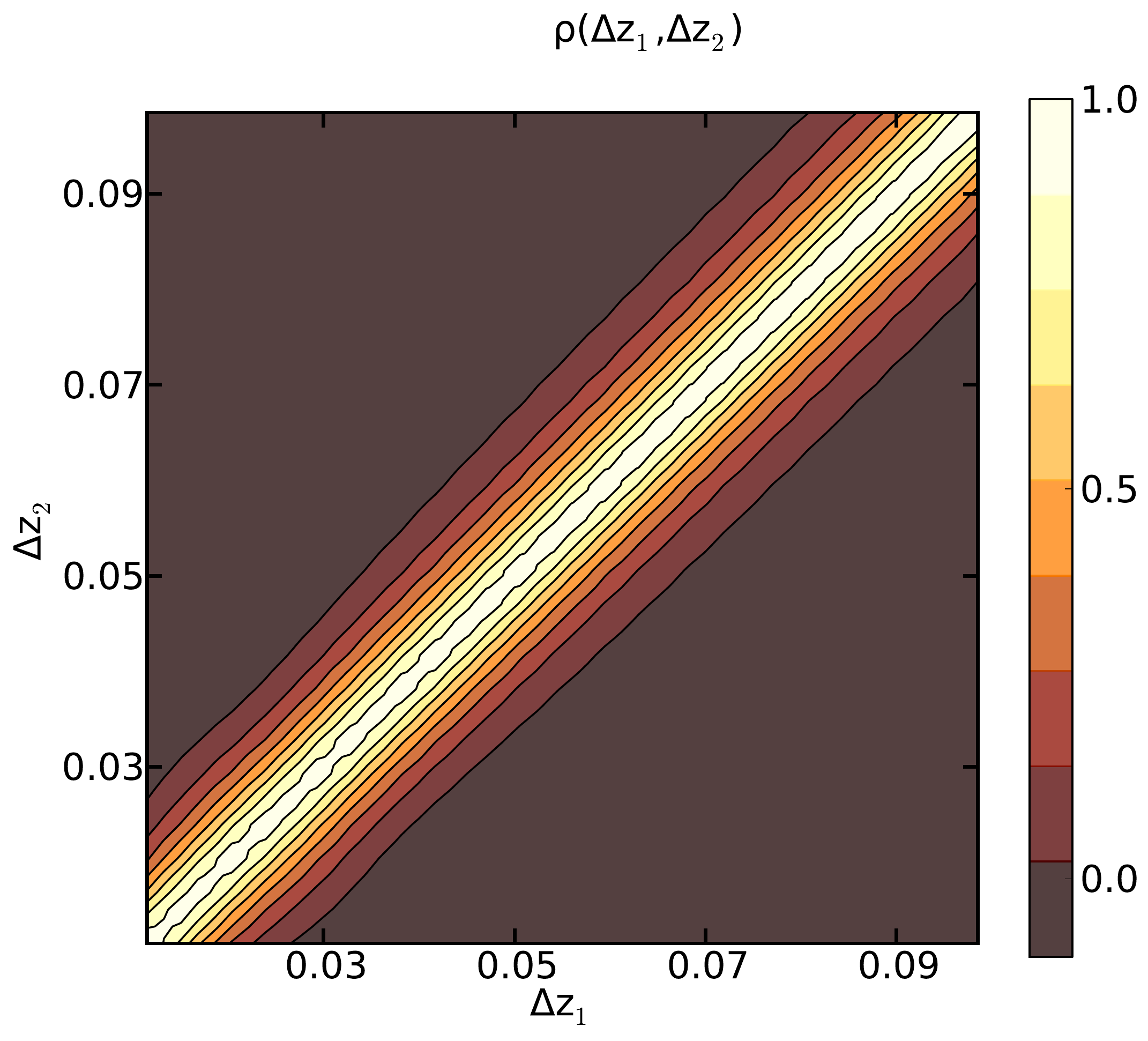}
    \includegraphics[width=0.285\textwidth]{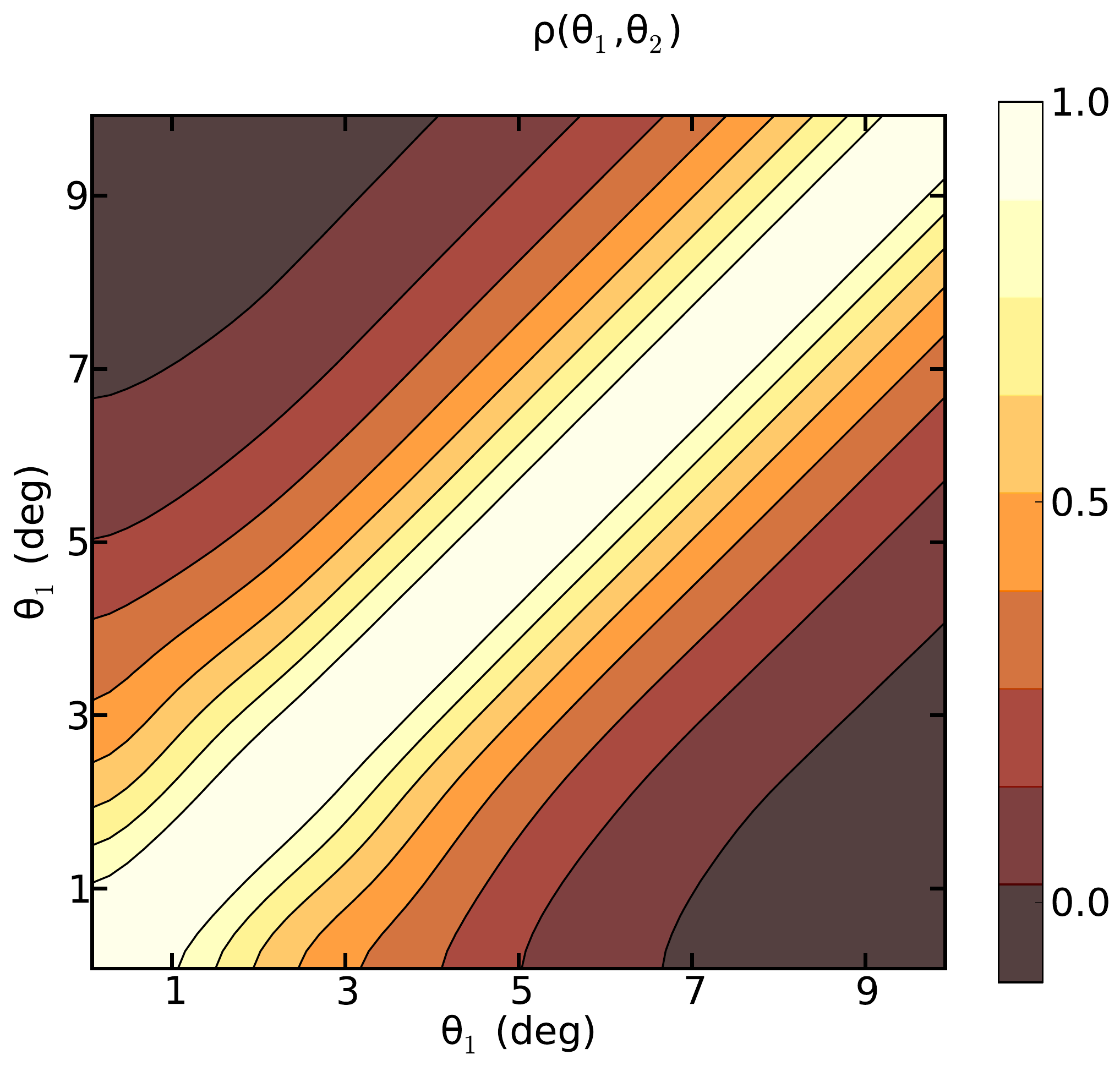}
    \includegraphics[width=0.30\textwidth]{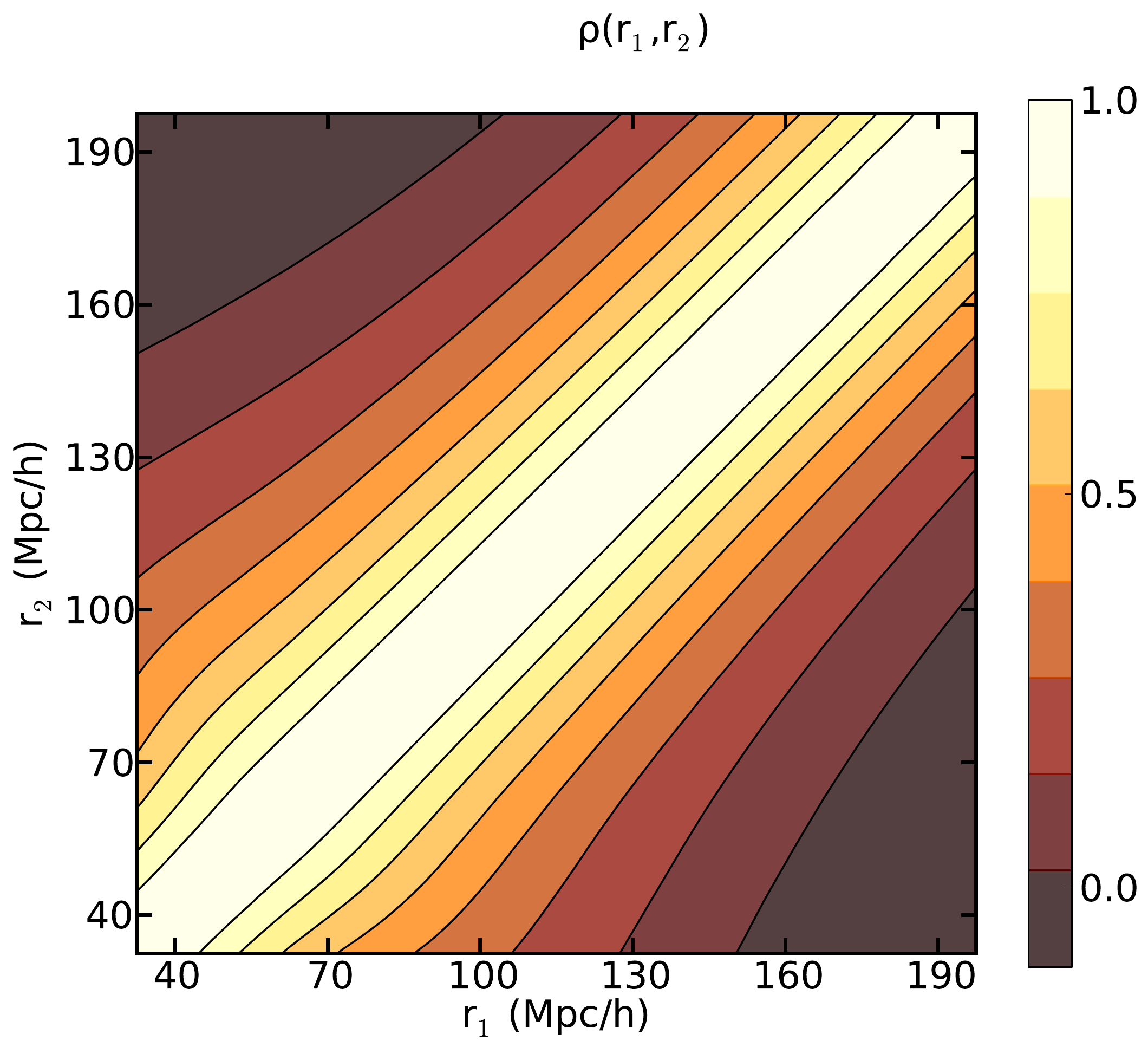}
  \end{tabular}
  \caption{Reduced Gaussian covariance matrices for the radial (left), angular (center) and
           monopole correlation functions. The covariances in the first and last cases were
           computed for a survey with 5000 square-degrees in $0.45 < z < 0.75$, while the
           angular case was computed for a bin $0.5 < z < 0.6$. In the three cases the axis
           ranges correspond to similar comoving scales.}
  \label{fig:covariances}
\end{figure*}

We have used a theoretical estimation of the covariance matrix for the 3-D correlation
function, which is then applied to the estimation of the covariance for the radial 
correlation function. Our approach is based on assuming Gaussian statistics for the overdensity
field, in analogy with the approach in \citet{2013MNRAS.431.2834X} for the monopole and with
the method described in \cite{2011MNRAS.414..329C} for the angular correlation function. We
have used the following convention for the power spectrum:
\begin{eqnarray}
\langle\delta_{\bm k}\delta_{{\bm k}^{\prime}}\rangle 
&= \delta^{D} (\bm{k}+\bm{k}^{\prime}) \left(P({\bf k})+\bar{n}^{-1}\right) \\
&= \frac{V}{(2 \pi)^3} \delta^{K}_{\bm{k}-\bm{k}^{\prime}} \left(P({\bf k})+\bar{n}^{-1}\right)
\end{eqnarray}
where $\delta^{D}$ and $\delta^{K}$ are the Dirac and Kronecker deltas, $V$ is the volume and 
we have taken into account the Poisson contribution to the total variance as the inverse of the
number density of sources $\bar{n}$.

The anisotropic power spectrum can be estimated from a given realization of $\delta_{\bf k}$ by
averaging over the symmetric azimuthal angle:
\begin{equation}
  \hat{P}(k_{\parallel},k_{\perp})\equiv\frac{1}{2\pi}\int_0^{2\pi}d\phi_k
  \left(\frac{(2\pi)^3}{V}|\delta_{\bf k}|^2-\frac{1}{\bar{n}}\right)
\end{equation}
    
Assuming that $\delta_{\bm k}$ is Gaussianly distributed we can obtain the covariance for
this estimator using Wick's theorem:
\begin{align}\nonumber
  C^P({\bf k}_1,{\bf k}_2)= & 
  \frac{4\pi^2}{V}\mathfrak{P}^2(k_{\parallel},k_{\perp})
  \frac{\delta^{\mathcal{D}}(k_{2,\perp}-k_{1,\perp})}{k_{1,\perp}}\\
  & \left[\delta^{\mathcal{D}}(k_{2,\parallel}-k_{1,\parallel})+
    \delta^{\mathcal{D}}(k_{2,\parallel}+k_{1,\parallel})\right],
\end{align}
where, as is done in \citet{2013MNRAS.431.2834X}, the effect of a non-homogeneous number
density $\bar{n}(z)$ has been taken into account by defining the volume-averaged
variance
\begin{align}\nonumber
  \mathfrak{P}^{-2}(k_{\parallel},k_{\perp})\equiv\int\frac{dV(z)}
       {P(k_{\parallel},k_{\perp})+\bar{n}^{-1}(z)}.
\end{align}
The anisotropic power spectrum can be related to the anisotropic correlation function
using the 0-order cylindrical Bessel function $J_0(x)$ through
\begin{equation}\nonumber
 \xi(\pi,\sigma)=\frac{1}{4\pi^2}\int_{-\infty}^{\infty}dk_{\parallel}e^{i\,k_{\parallel}\pi}
 \int_0^{\infty}dk_{\perp}k_{\perp}J_0(k_{\perp}\sigma)P(k_{\parallel},k_{\perp}),
\end{equation}
And thus the covariance matrix for $\xi$ can be calculated as
\begin{align}\nonumber
  C^{\xi}({\bf r}_1,{\bf r}_2)\equiv&\frac{1}{\pi^2}
  \int_0^{\infty}dk_{\parallel}\,\bar{\cos}(k_{\parallel}\pi_1)\,\bar{\cos}(k_{\parallel}\pi_2)\\
  &\int_0^{\infty}dk_{\perp}k_{\perp}\bar{J}_0(k_{\perp}\sigma_1)\,
  \bar{J}_0(k_{\perp}\sigma_2)\mathfrak{P}^2(k_{\parallel},k_{\perp}).
\end{align}
Here we have taken into account the finiteness of the bins $\Delta\pi$, $\Delta\sigma$ by
defining the regularized functions $\bar{\cos}$ and $\bar{J}_0$:
\begin{align}\nonumber
  &\bar{\cos}(k_{\parallel}\pi)\equiv\frac{\sin(x_2)-\sin(x_1)}{x_2-x_1},
   \hspace{6pt} (x_{1,2}\equiv k_{\parallel}(\pi\pm\Delta\pi)) \\\nonumber
  &\bar{J}_0(k_{\perp}\sigma)\equiv2\,\frac{x_2\,J_1(x_2)-x_1\,J_1(x_1)}{x_2^2-x_1^2},
   \hspace{6pt} (x_{1,2}\equiv k_{\perp}(\sigma\pm\Delta\sigma))
\end{align}

From this result it is straightforward to compute the covariance matrix for the 
radial correlation function ($\sigma_{1,2}=0$) as
\begin{equation}\label{eq:gcv_radial}
  C^{\xi}_{\parallel}(\pi_1,\pi_2)\equiv\frac{1}{\pi^2}
  \int_0^{\infty}dk_{\parallel}\,\bar{\cos}(k_{\parallel}\pi_1)\,
  \bar{\cos}(k_{\parallel}\pi_2)\,
  \left[\mathfrak{P}_{\parallel}(k_{\parallel},\Delta \sigma)\right]^2,
\end{equation}
where we have defined the projected $k$-space variance
\begin{equation}\label{eq:kspace_variance}
  \left[\mathfrak{P}_{\parallel}(k_{\parallel},\Delta\sigma)\right]^2\equiv
  \int^{\infty}_{0}dk_{\perp}\,k_{\perp}
  \left[2\frac{J_1(k_{\perp}\,\Delta\sigma)}{k_{\perp}\,\Delta\sigma}\right]^2
  \mathfrak{P}^2(k_{\parallel},k_{\perp}).
\end{equation}
Here the radial coordinate $\pi$ is related to the redshift separation through
$\Delta z \equiv \pi\,H(z)$, and the transverse width $\Delta\sigma$ is related
to the angular pixel resolution through 
$\Delta \theta \equiv \Delta\sigma/r(z)$. As can be verified on closer inspection 
of equation \ref{eq:kspace_variance}, the covariance grows dramatically as we 
decrease the pixel size. This is due to the fact that the number of galaxy pairs 
drops and the errors become shot-noise dominated.

Throughout this work, Gaussian estimations for the covariance matrices were used.
Figure \ref{fig:covariances} shows the Gaussian predictions for the reduced covariance
matrix $\rho_{ij}\equiv C_{ij}/\sqrt{C_{ii}\,C_{jj}}$ of the radial correlation
function (left panel), the angular correlation function (central panel) and the 
monopole (right panel). The axis ranges correspond to similar comoving scales in the
three cases. Notice that, while the errors on the radial correlation function are
almost diagonal, they are correlated over a large range of scales in the angular
(transverse) case. For the monopole, the errors are also correlated over
a large number of bins, which is a sensible result, since the monopole corresponds
to an angular average over two transverse and one longitudinal directions.

\section{Angular BAO results for the Combination}
\label{sec:Angular_BAO}

\begin{figure}
  \centering
  \includegraphics[width=0.49\textwidth]{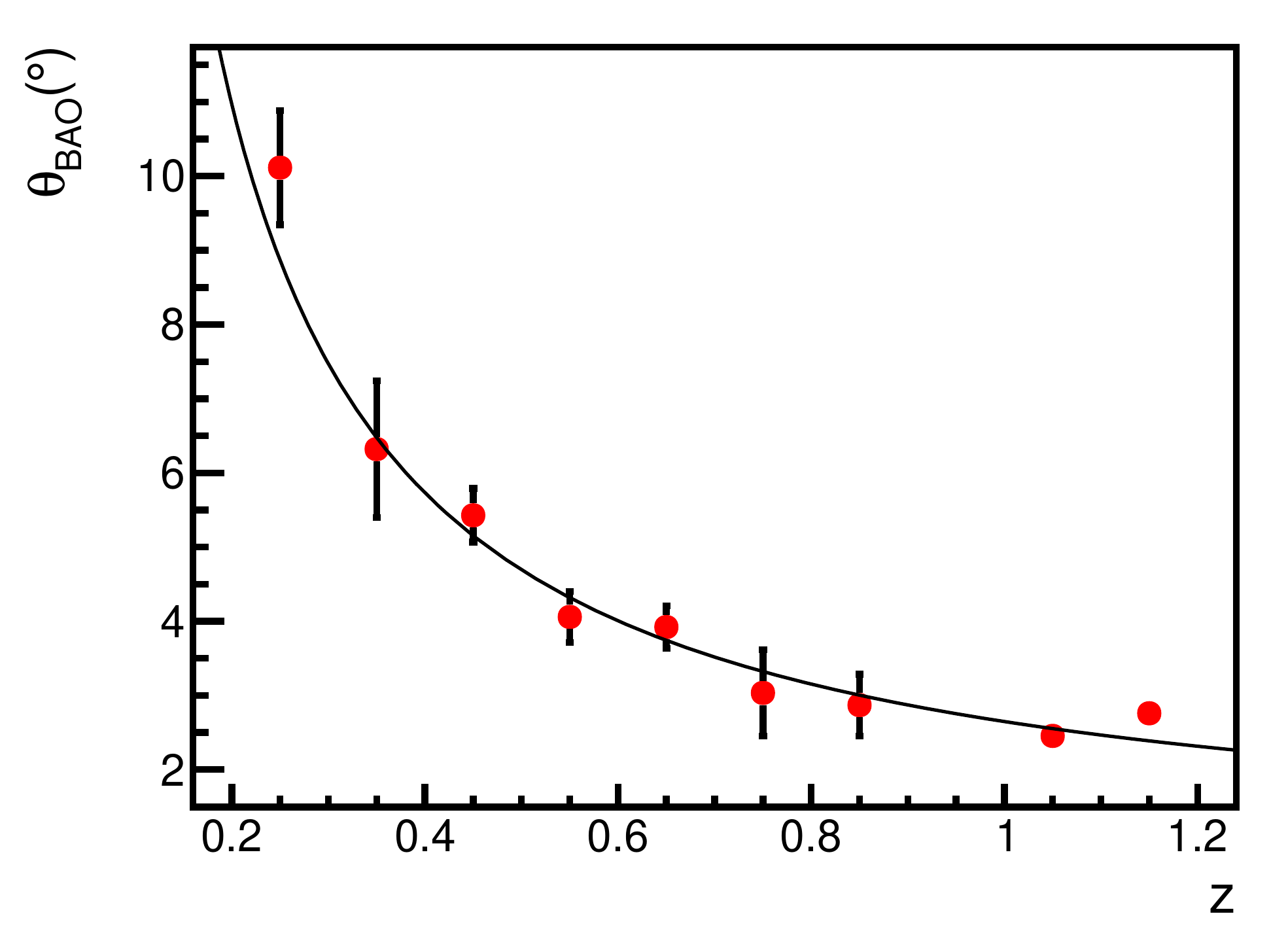}
  \caption{Measured angular BAO scale as a function of the redshift in the 
           MICE simulation using the method described in 
           {\protect \cite{2011MNRAS.411..277S}}. Dots are the measured
           values of $\theta_{BAO}$ and the solid line is the prediction for 
           the cosmology of the simulation.}
  \label{fig:thetabao_vs_z}
\end{figure}

The angular BAO measurements we have used for the results presented in the text
have been obtained applying the method described in \cite{2011MNRAS.411..277S} to
the cosmological simulation used in this paper. We have used 10 redshift bins of
width 0.1, starting at redshift 0.2 up to redshift 1.2. We find statistically 
significant results in 9 of them. Results are presented in 
Figure~\ref{fig:thetabao_vs_z}. These are the results we combine with the radial
BAO in order to obtain the cosmological constraints presented in the text. It is
important to notice that the redshifts for this galaxy sample are spectroscopic
and consequently, the systematic error associated to the photometric redshift
that is quoted on \cite{2011MNRAS.411..277S} does not affect these measurements.

\end{document}